\documentclass[3p,authoryear, final]{elsarticle}

\usepackage{hyperref}

\usepackage{setspace}
\usepackage{graphicx}
\usepackage[cmex10]{amsmath}
\usepackage{array}
\usepackage{subfigure}
\usepackage{url}
\usepackage{framed, color}
\usepackage{tikz}
\usepackage{units}
\usepackage{natbib}
\usetikzlibrary{backgrounds,fit,shapes.multipart,plotmarks,patterns,shapes,arrows}
\usepackage{xspace}

\definecolor{be_green}{rgb}{0.549019607843137, 0.705882352941177, 0.0392156862745098} 
\definecolor{be_blue}{rgb}{0.235294117647059, 0.705882352941177, 0.862745098039216} 
\definecolor{blue2}{rgb}{0.2, 0.55, 0.59} 

\tikzstyle{block} = [rectangle, draw, fill=gray!20, 
    text width=2.5cm, text centered, rounded corners, minimum height=0.8cm]
\tikzstyle{textblock} = [rectangle, draw, fill=gray!20, 
    text width=2.3cm, text centered, rounded corners, minimum height=0.5cm]    
\tikzstyle{textblockSmall} = [rectangle, draw, fill=gray!20, 
    text width=1.8cm, text centered, rounded corners, minimum height=0.5cm] 
\tikzstyle{textblockWide} = [rectangle, draw, fill=gray!20, 
    text width=6cm, text centered, rounded corners, minimum height=0.5cm]  
         
\tikzstyle{line} = [draw, -latex']

\newcommand{\ie}{i.e.}

\newcommand{\eg}{e.g.}

\renewcommand{\d}[1]{{\mbox{\boldmath$#1$}}}

\newcommand{\m}[1]{{\mbox{{\fontencoding{T1}\sffamily\slshape{#1\/}}}}}

\newcommand{\ical}[1]{{\mbox{\usefont{OT1}{pzc}{m}{it}{#1}}}}

\newcommand{\STAR}{\textsc{STAR{}}}
\newcommand{\sgdr}{\textsc{SGDR}}
\newcommand{\thresh}{\textsc{ICE1}}
\newcommand{\wtp}{\textsc{W3POM}}
\newcommand{\itr}{\textsc{ITR}}
\newcommand{\ales}{\textsc{ALES}}

\newcommand{\STARnt}{$N_{\text{\STAR{}}}$}
\newcommand{\sgdrnt}{$N_{\text{\sgdr{}}}$}
\newcommand{\itrnt}{$N_{\text{\itr{}}}$}
\newcommand{\threshnt}{$N_{\text{\thresh{}}}$}
\newcommand{\wtpnt}{$N_{\text{\wtp{}}}$}
\newcommand{\alesnt}{$N_{\text{\ales{}}}$}

\newcommand{\STARrms}{$\sigma_{\text{\STAR{}}}$}
\newcommand{\sgdrrms}{$\sigma_{\text{\sgdr{}}}$}
\newcommand{\itrrms}{$\sigma_{\text{\itr{}}}$}
\newcommand{\threshrms}{$\sigma_{\text{\thresh{}}}$}
\newcommand{\wtprms}{$\sigma_{\text{\wtp{}}}$}
\newcommand{\alesrms}{$\sigma_{\text{\ales{}}}$}

\begin{document}


\begin{frontmatter}

\journal{Remote Sensing of Environment}

\title{STAR: Spatio-Temporal Altimeter Waveform Retracking using Sparse~Representation and Conditional Random Fields}

\author{Ribana Roscher, Bernd Uebbing, J\"urgen Kusche}

\address{Institute of Geodesy and Geoinformation, University of Bonn, Nu\ss allee 15+17, 53115 Bonn, Germany}



\begin{abstract}
Satellite radar altimetry is one of the most powerful techniques for measuring
sea surface height variations, with applications ranging from operational oceanography to climate research. 
Over open oceans, altimeter return waveforms generally correspond to the Brown
model, and by inversion, estimated shape parameters provide mean surface height and wind speed. 
However, in coastal areas or over inland waters, the waveform shape is
often distorted by land influence, resulting in peaks or fast decaying trailing
edges. As a result, derived sea surface heights are then less accurate and
waveforms need to be reprocessed by sophisticated algorithms. To this end, 
this work suggests a novel Spatio-Temporal Altimetry Retracking (STAR) technique. 
We show that STAR enables the derivation of sea surface heights over the open ocean
as well as over coastal regions of at least the same quality as compared to
existing retracking methods, but for a larger number of cycles and thus
retaining more useful data. Novel elements of our method
are (a) integrating information from spatially and temporally
neighboring waveforms through a conditional random field approach,  (b)
sub-waveform detection, where relevant sub-waveforms are separated from
corrupted or non-relevant parts through a sparse representation approach, and
(c) identifying the final best set of sea surfaces heights from multiple
likely heights using Dijkstra's algorithm. We apply STAR to
data from the Jason-1, Jason-2 and Envisat missions for study sites in the Gulf of
Trieste, Italy and in the coastal region of the Ganges-Brahmaputra-Meghna estuary,
Bangladesh. We compare to several established and recent retracking methods, as well as to tide gauge data. Our experiments suggest that the obtained sea surface heights are significantly less affected by outliers when compared to results obtained by other approaches. 
\end{abstract}

\begin{keyword}
coastal \sep oceans \sep altimetry \sep retracking \sep sea surface heights \sep conditional random fields \sep sparse representation
\end{keyword}

\end{frontmatter}


\linespread{1.5}
\section{Introduction}
For several decades, radar altimetry is routinely being used for monitoring
sea surface height (SSH) variations. Observed SSHs play a key role in several
applications, ranging from operational oceanography \citep{Chelton2001} and
tidal modeling \citep{Wang2004, Savcenko2008} to gravity estimation
\citep{Hwang1998}, and they serve as important indicators in climate
research. Recently, radar altimetry in coastal zones \citep{Gommenginger2011}
and for inland water bodies \citep{Birkett2010} has become a topic of
increasing interest. However, in both applications one needs to mitigate the
potentially significant land influence on the altimeter return signal.\\

The altimeter instrument on-board a satellite emits a spherically propagating,
nadir-directed radar pulse, which is reflected at the
surface. Range information can then be inferred from the two-way travel
time \citep{Fu2001}. In addition, the returned signal energy is measured over time,
forming an altimeter waveform. It can be shown that over an ideal surface, the return waveform
corresponds to the theoretical Brown model \citep{Brown1977} and the estimated
shape parameters of this model provide information on mean SSH and significant wave height (SWH), while the amplitude strength of the reflected radar pulse can be used to derive wind speed. On board the satellite, the waveform signal is sampled at discrete epochs with a spacing of about \unit[3]{ns} of two-way travel time, which are generally referred to as range gates \citep{Chelton1989}. Altimeter measurements do not refer to an individual point directly below the satellite, but rather to a footprint with a diameter of several kilometers, depending on SWH and the altitude of the altimetry mission. 

\begin{figure}[ht]
\centering
\includegraphics[width=0.6\columnwidth]{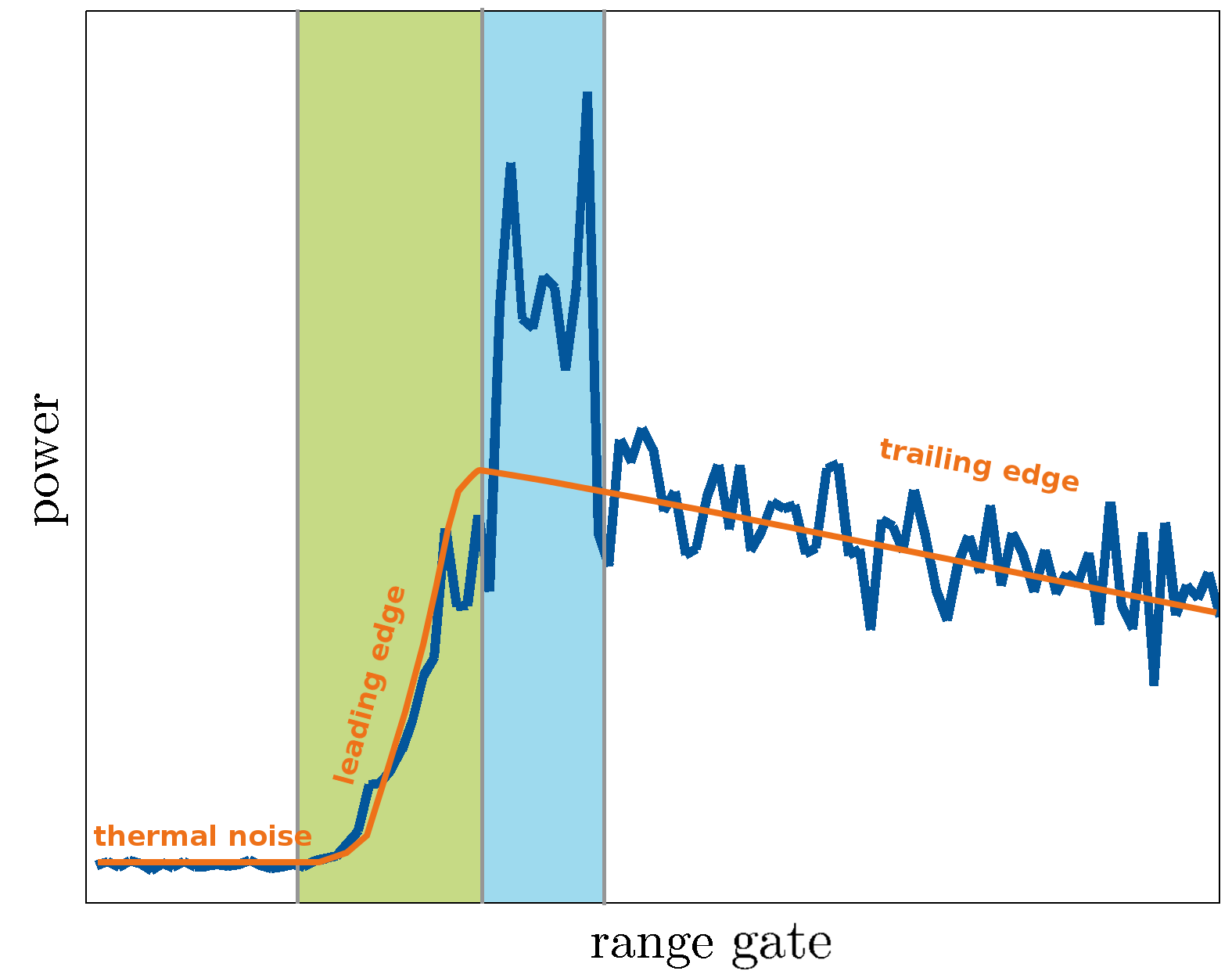}
\caption{Waveform with disturbing peak caused by land influences (colored in blue). The relevant part for sea surface height, determined by sub-waveform detection, is illustrated in green. A theoretical waveform model is depicted in orange.}
\label{fig:relevant}
\end{figure}

As illustrated in Fig. \ref{fig:relevant}, the return waveform over the open ocean consists of three main parts;
first, before any return energy from the radar pulse is measured, the waveform
contains only thermal noise which is present in all radar systems. As soon as
the front of the radar pulse hits the wave crests, the altimeter footprint is
defined by a single point and the measured return energy begins to
rise. Afterwards, more of the pulse illuminates the surface around the initial
point and the footprint becomes a growing circle, which corresponds to rapidly
increasing signal energy in the measured altimeter waveform. The leading edge
of the altimeter return waveform is defined between the first energy return
from wave crests and the return energy after the radar pulse has reached the
wave troughs (Fig. \ref{fig:relevant}). At this point. the area of the
footprint circle reaches its maximum, which is defined as the pulse--limited
footprint \citep[PLF, ][]{Chelton1989}. Afterwards, the circle transforms into
an annulus with increasing inner and outer radii, but with a fixed
illuminated area. The corresponding signal energy measured outside of the PLF is referred to as trailing edge of the measured waveform (Fig. \ref{fig:relevant}). The slope of the trailing edge can be utilized to derive information on the off-nadir attitude of the altimeter satellite. \\

The $50$\%-point on the leading edge corresponds to the mean sea level between
wave crests and wave troughs, and thus represents the reference point for the
range measurement. An algorithm on board of the satellite tries to position this
point inside a pre-defined range window at a fixed tracking range gate
\citep[31 for Jason-2, ][]{Quartly2001}. This range window consists of a fixed
number of range gates, covering about \unit[50]{m} depending on the satellite
mission, and it is positioned by the onboard tracker based on prior information on the
range. However, positioning is not always perfect and the $50$\%-point is
not located exactly at the tracking gate. Consequently, this requires a
ground-based reprocessing of the altimeter waveforms transmitted back to
Earth, a procedure which is called retracking. Over the open ocean, the shape
of the waveform will be close to the theoretical Brown model with the $50$\%-point
being only slightly shifted from the tracking gate position, and this can be
easily corrected using an ocean model retracker \citep{Brown1977, Hayne1980,
  Deng2003}. For ice surfaces where the waveform signal often contains two
leading edges due to the radar signal partly penetrating through the upper
snow layer, specialized retracking algorithms have been developed
\citep{Martin1983}. However, in coastal areas the waveform shape is typically
disturbed by land influences in the altimeter footprint, resulting in peaks or fast decaying
trailing edges.

These deviations of coastal waveforms from the Brown model lead conventional
ocean retrackers to generate diverging or strongly biased estimates of SSH, as
land-induced peaks propagate along the trailing edge towards the leading edge
while the altimeter ground track approaches the coast \citep{Lee2010}.
In order to mitigate the land influences on the waveform shape, various
tailored approaches have been proposed. As an example for methods that seek to
model the entire waveform, \citep{Halimi2013} combined a 3-parameter Brown
ocean model with a modeled asymmetric peak to account for land influences. A
different approach for dealing with the influence of peaks on the retracked
estimates is to first partition the waveform in a pre-processing step; \ie{}
to identify relevant parts of the waveform, such as the leading edge, but also
possible peaks. For example, \citep{Hwang2006} first identify relevant
sub-waveforms and then apply a threshold retracking algorithm to each of the
sub-waveforms, which leads to multiple equally likely SSH estimates at each location
from which the final estimate is chosen based on comparison to a-priori
height information. In this way, peaks that appear outside of the
relevant sub-waveforms are ignored. Recently, for retracking SSHs over
inland water bodies, \citep{Uebbing2015}
combined the sub-waveform approach from \citep{Hwang2006} and the waveform
model from \citep{Halimi2013} to suppress land-induced peaks on the trailing
edge, but also to account for possible peaks close to the leading edge of the
waveform. This could be shown to lead to improved lake heights compared to conventional methods. 
In a different approach, \citep{Passaro2014} suggested a two-step procedure, similar to a previously published approach by \citet{Sandwell2005}, where in the first step all 3 parameters (amplitude, range and SWH) are
estimated. In the second iteration they fixed the SWH to a mean value
derived from the first step and re-estimated the amplitude and range
correction, since SWH estimations are strongly correlated to the range
correction. This leads to improved SSHs closer to the coast.

Here, we introduce a novel method for the analysis of sea surface heights from altimetric waveforms, which will utilize spatial information from neighboring range gates within one waveform, as well as temporal information from neighboring waveforms along the altimeter track. This Spatio-Temporal Altimetry Retracker (STAR) can be applied to altimetry data over the open ocean, as well as in coastal areas. Our contributions are twofold: First, our analysis includes a novel sub-waveform detection scheme, which to our knowledge for the first time integrates spatial as well as temporal information. This differs from the conventional sub-waveform detection algorithm \citep{Hwang2006} in that we partition the entire waveform into separate sub-waveforms, instead of identifying possible, disjointed leading edges. Second, in order to be largely independent of the choice of tuning (or 'hyper') parameters within the sub-waveform detection scheme, we derive multiple sub-waveform partitionings by varying the weight $w$ between unary and binary terms of the conditional random field. This leads to a range of partitionings of the entire waveform, and subsequently to a point cloud of equally likely SSHs at each measurement position, each of which is estimated using a 3-parameter ocean model \citep{Halimi2013}.
We then employ Dijkstra algorithm \citep{Dijkstra1959} to find reasonably smooth SSHs, without resorting to fitting. 

Our sub-waveform detection scheme uses a sparse representation (SR) approach, where the return power at all range gates within one particular sub-waveform is modeled by a weighted linear combination of a single common set of basis waveforms, which are derived from synthetic Brown waveforms. The concept of SR has been applied to many areas of signal analysis \citep{Wright2010}, but this study appears to be the first which uses it on radar altimetry.
 
SSHs and other sea surface conditions such as wave height are neither independent along tracks, nor between neighboring tracks (\cite{Sandwell1997}).
Spatial information has been used in the analysis, for example, by \cite{Maus1998} through simultaneously processing of a sequence of waveforms for tracking of travel times, or \cite{Halimi2016a} for a smooth estimation of altimetric parameters.
This means, the integration of spatial information can be carried out in different parts of the analysis.
The latter two approaches, for example, integrate spatial information about neighboring waveform to develop improved estimation algorithms for retracking.  
Here, we integrate spatio-temporal information by means of a conditional random field \citep[CRF, ][]{Lafferty2001}: to this end, we introduce spatial relations between the return power at range gates of temporally neighboring waveforms, \ie{} neighboring waveforms within one pass and cycle, and relations between the range gate power within a single waveform. In this way, range gates which are relevant for SSH estimation can be distinguished from corrupted or non-relevant waveform parts, since they are represented through a different linear combination of basis sub-waveforms (see Fig. \ref{fig:relevant}). 
In contrast to \citet{Halimi2016a} where the conditional random field is used as part of an algorithm to enforce a smooth estimation of the retracking parameters, we propose to use the conditional random field at the sub-waveform detection step.  
Subsequently, any retracking method can be applied to the identified individual sub-waveforms for deriving the SSH, thus effectively ignoring disturbing signals outside the selected sub-waveform. This means, our approach could be transferred in future to the analysis of Delay-Doppler altimetry.

This paper is organized as follows:
First, we introduce the altimetry data that we used for validating our method, as well as our two study sites in Sec. \ref{sec:dataMain}.
We describe the sub-waveform detection with SR and the integration of spatio-temporal knowledge by employing a CRF in Sec. \ref{sec:sub}.
Moreover, the estimation of single SSHs and the determination of the best set of SSHs from multiple heights using Dijkstra's algorithm is explained in Sec. \ref{sec:sshe}. In Sec. \ref{sec:experiments}, we compare the performance of our sub-waveform detection against an existing method and we evaluate the proposed analysis framework by means of analysing Jason-1, Jason-2 and Envisat waveforms in the coastal regions of the Gulf of Trieste, and in the northern Bay of Bengal in the coastal waters of Bangladesh.
Sec. \ref{sec:conclusion} concludes the paper.

\section{Data and Study Sites}
\label{sec:dataMain}
\subsection{Data}
\label{sec:data}
We apply our retracking method to Jason-2 Sensor Geophysical Data Records (SGDR) of the Jason-2 mission, as well as to SGDRs from the Jason-1 and European Environmental Satellite (Envisat) missions. 

The Ocean Surface Topography Mission (OSTM)/ Jason-2 was launched in mid of 2008 succeeding the Jason-1 mission on the same orbit. The satellite flies in a near circular $\sim$10 day repeat orbit with an altitude of \unit[1336]{km} and inclination of \unit[66]{$^\circ$} and a separation of the groundtracks equal to \unit[315]{km} at the equator. The main instrument is the Poseidon-3 altimeter which emits radar pulses in the Ku-Band (\unit[13.575]{GHz}/\unit[2.21]{cm}) and C-Band (\unit[5.3]{GHz}/\unit[5.08]{cm}) \citep{Aviso2015}. Additional instruments are a microwave radiometer used to derive the wet troposphere correction, as well as GPS and DORIS systems for precise orbit determination \citep{Rosmorduc2011}. The Jason-2 SGDRs have been obtained from the Archiving, Validation and Interpretation of Satellite Oceanographic (AVISO) team which is part of Centre National d'Etudes Spatiales (CNES). The SGDRs are sorted by pass and cycle, including 254 passes per cycle or $\sim$10 day repeat orbit and we utilize data of passes 053 and 196 from the beginning of the mission in July 2008 (cycle 0) until the end of 2014 (cycle 239).

The composition of the Jason-1 mission is very similar to Jason-2. It was launched in December, 2001 as a successor to the Topex/Poseidon mission. After the launch of Jason-2 in June, 2008 both satellites flew on the same orbit in close distance to allow intercalibration of the satellite missions. After 6 months, Jason-1 was moved to an interleaved orbit located in the middle between the nominal orbit to increase the spatial resolution of the combined Jason-1 and Jason-2 data until January, 2012. Afterwards the satellite was moved to a drifting geodetic orbit and passivated and decommissioned in July, 2013 after losing contact. We will use data from the interleaved period from February, 2009 (cycle 263) until January, 2012 (cycle 370). The Jason-1 SGDRs for the interleaved orbit are acquired from the Physical Oceanography Distributed Active Archive Center (PO.DAAC, \url{ftp://podaac.jpl.nasa.gov/allData/coastal_alt/L2/ALES/jason-2/}) operated by the Jet Propulsion Laboratory (JPL) which is part of the National Aeronautics and Space Administration (NASA). 

The Envisat satellite was launched in March, 2002 succeeding the ERS-2 mission in the same orbit by the European Space Agency (ESA). The orbit is a 35-day repeat orbit with an altitude of \unit[800]{km}, about an inclination of \unit[98.55]{$^\circ$} which allows the satellite to cover higher latitude regions compared to \eg{} Jason-2 with a higher spatial resolution (\unit[80]{km} separation at the equator) at the cost of a longer repeat period of 35 days. The satellite carries a total of 10 instruments of which the DORIS positioning system, the microwave radiometer and the Radar Altimeter 2 (RA2) altimeter instrument are of most importance to us \citep{Rosmorduc2011}. The RA2 altimeter is a dual frequency altimeter emitting radar pulses in Ku-Band (\unit[13.575]{GHz}/\unit[2.21]{cm}) and S-Band (\unit[3.2]{GHz}/\unit[9.37]{cm}) \citep{envisatHB2007}. The Envisat SGDR data are provided by ESA (\url{https://earth.esa.int/}). We utilize data from June, 2002 (cycle 7) to September, 2010 (cycle 93).

We extract the \unit[20]{Hz} (\unit[18]{Hz} for Envisat) tracker range, altitude and waveforms which are needed during the retracking algorithm. Additionally, \unit[1]{Hz} atmospheric model corrections for the dry and wet troposphere, as well as the ionosphere are extracted from the SGDR data and linearly interpolated to the high rate positions.

For validation of the retracked coastal SSHs, tide gauge data with hourly resolution from the University of Hawaii Sea Level Center (UHSLC) are used. The hourly data are uncorrected with respect to tidal and inverse barometric effects and we do not apply these corrections to neither the altimetry data, nor the tide gauge data to remove the influence of these corrections on the validation. The tide gauge data for Trieste, Italy is available from June, 2009 to December 2015 and for the tide gauge station in Chittagong, Bangladesh we have data from July, 2007 til December 2015.
Additionally, we utilize openly available GDR datasets for the Jason-2 (cycles 1 to 239) and Envisat (cycles 7 to 93) mission that include the ALES retracked ranges for comparison which are distributed by PO.DAAC, JPL. 

For the validation of significant wave height and wind speed, we utilize model data from the ERA-Interim reanalysis \citep{Dee2011} which is distributed by the European Centre for Medium-Range Weather Forecast (ECMWF) and interpolated to the altimetry track.

\subsection{Study Sites}
For investigating the quality of our proposed STAR algorithm, we selected two study sites located in the Gulf of Triest and in the coastal regions of Bangladesh. The sites include varying conditions, including shallow coastal waters, open ocean areas, temporally submerged sand banks and transition zones between river estuaries and the ocean. 

\begin{figure}[ht]
	\centering
	\subfigure[Triest TG at the Adriatic Sea and the groundtracks of Jason-2 pass 196, as well as Jason-1 interleaved orbit, pass 161.]{
	\includegraphics[height=0.4\columnwidth]{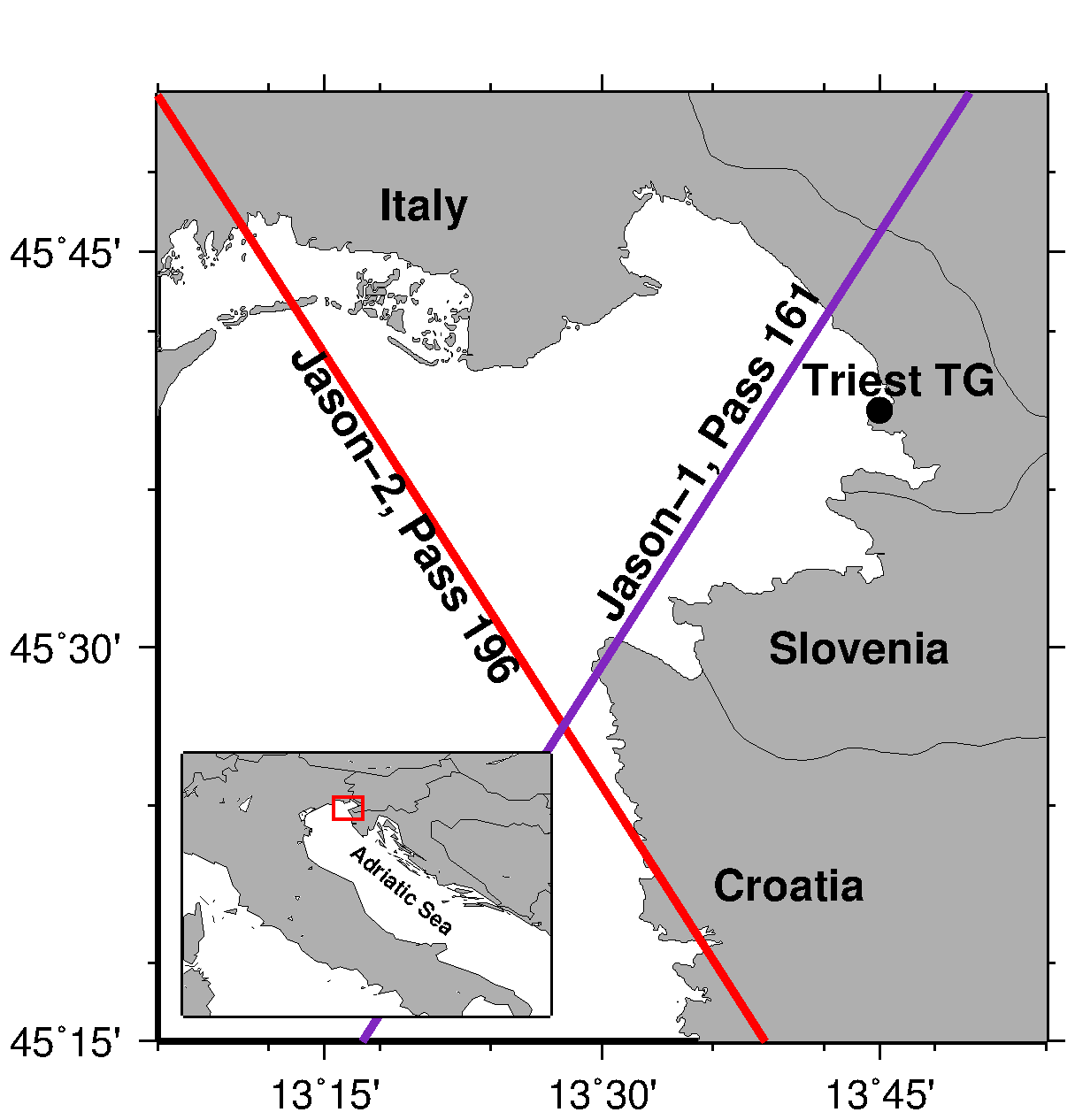}}\quad
	\subfigure[Chittagong TG at the Bay of Bengal and the groundtracks of Jason-2 nominal orbit, pass 053, as well as Envisat, pass 0352.]{
	\includegraphics[height=0.4\columnwidth]{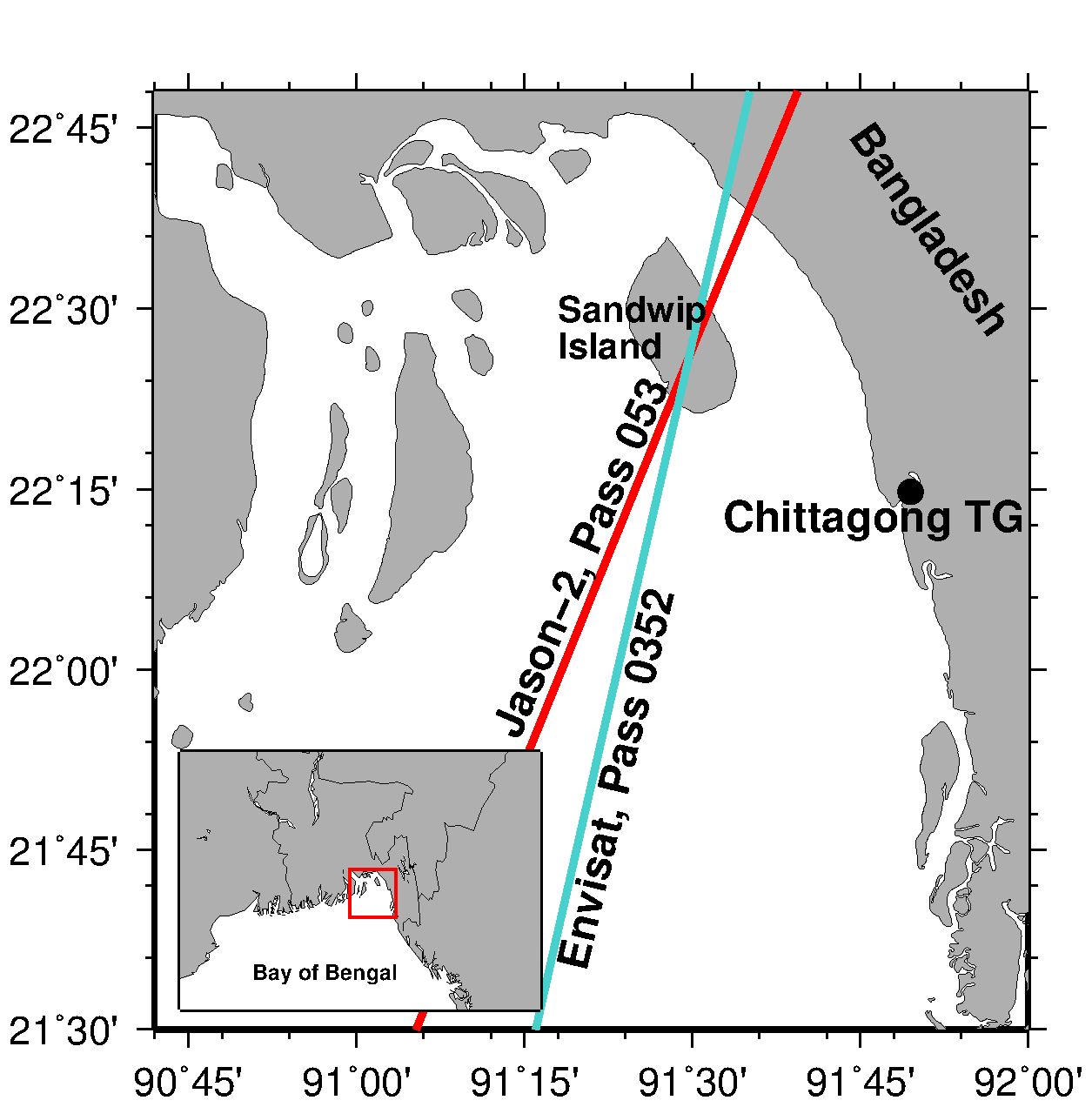}}
  	\caption{Location of the altimetry tracks and the tide gauges used for validation of the retracked SSHs.}
\label{fig:study}
\end{figure}

\subsubsection{Triest}
The first study site is located in the northern Adriatic Sea in the north-eastern part of the Gulf of Venice and includes the Gulf of Triest. The descending nominal orbit pass 196 of the Jason-2 mission crosses the study area from north-west to the south-east (Fig. \ref{fig:study}(a)). It crosses from the Italian mainland to ocean close to the city of Marano Lagunare at approx \unit[45.76]{$^\circ$N} and covers about \unit[5.9]{km} of the Laguna di Marano before there is a short ocean-land-ocean transition over the Isola di Sant' Andrea at $\sim$\unit[45.71]{$^\circ$N}. Then, the track covers about \unit[45.5]{km} of open ocean in the Gulf of Venice before transitioning to the Croatian mainland at $\sim$\unit[45.36]{$^\circ$N}. For the last \unit[5]{km} from $\sim$\unit[45.4]{$^\circ$N} until the track is over the Croatian mainland, it runs very close to the Croatian coast with a distance of less than \unit[2]{km}.

Furthermore, we utilize data from the ascending Jason-1 pass 161 of the interleaved orbit (Fig. \ref{fig:study}(a)). It crosses our study area from the south-west to the north-east. The first \unit[37]{km} are located over the open ocean and the groundtrack covers the Croatian mainland from $\sim$\unit[45.48]{$^\circ$N} to \unit[45.50]{$^\circ$N}. Then the track covers open water until it reaches the Italian mainland at $\sim$ \unit[45.71]{$^\circ$N}. At about \unit[45.54]{$^\circ$N} the track is located less than \unit[2]{km} away from the Slovenian mainland.

The tide gauge station Triest is located at $\sim$ \unit[13.75]{$^\circ$E} and \unit[45.65]{$^\circ$N} in the harbor of the city of Triest in the Gulf of Triest. The distance of the tide gauge station to the Jason-2 groundtrack is about \unit[31]{km} at the closest point.

\subsubsection{Bangladesh}
The second study site is located in the northern Bay of Bengal in a region right off the coast of Bangladesh. The region is covered by the ascending Jason-2 nominal orbit pass 053 which crosses the study area from the south-west to the north-east (Fig. \ref{fig:study}(b)). The track coverage starts over the open ocean parts of the Bay of Bengal and reaches Sandwip Island after $\sim$ \unit[95]{km} at $\sim$ \unit[22.41]{$^\circ N$}. From $\sim$ \unit[22.53]{$^\circ$N} to \unit[22.63]{$^\circ$N} the track again covers a strip of about \unit[13]{km} of open water related to the estuary of the Ganges-Brahmaputra-Meghna Delta (GBMD) before it reaches the Bangladesh mainland. Between latitudes of $\sim$ \unit[22.25]{$^\circ$N} and \unit[22.35]{$^\circ$N} there are some sand banks located along the track which are submerged during high tide, but not during low tide. 

Additionally, the descending pass 0352 of the Envisat mission runs almost parallel to the Jason-2 track and crosses the study site from the north-east to the south-west. It first reaches open water related to the GBMD at $\sim$\unit[22.67]{$^\circ$N} and after about \unit[14]{km} the track reaches Sandwip Island at $\sim$ \unit[22.55]{$^\circ$N}. At \unit[22.40]{$^\circ$N}, the track transitions back to open water and afterwards covers the remaining \unit[104]{km} of open ocean in our study area. Again, the sandbanks mentioned above might influence the data acquisition during low tide.

The Chittagong tide gauge station is located at $\sim$ \unit[91.83]{$^\circ$E} and \unit[22.25]{$^\circ$N} in the Chittagong harbor. The distance to the Jason-2 groundtrack is about \unit[38]{km} at the closest point.


\section{Sub-waveform Detection}
\label{sec:sub}
\subsection{Notation}
Let us consider $L$ consecutive waveforms collected along a cycle, each of which contains return energy for $G$ range gates. 
We arrange these waveforms in $\d x_l$, $l=1,\ldots,L$.
Each waveform can be represented through a set of overlapping windowed waveforms $\d \xi_{l,g} \in \d x_l$ with windows centered at $\xi_{l,g}$ and comprising $N_{\xi}$ neighboring range gates, as illustrated in Fig. \ref{fig:framework} for a single echo.
Within this framework, sub-waveform detection means identifying $K$ sub-waveforms in each $\d x_l$ by deciding for each range gate on indices $\hat{\d y}_{l,g}$, which define the best-fitting sparse representation (SR)-based models.
The aim of the proposed approach is to optimally detect these sub-waveforms, where the number of models per waveform is unknown and needs to be determined during the detection process.

\subsection{Detection Framework}
\label{sec:framework}
\begin{figure}[ht]
\centering
\includegraphics[width=1\textwidth]{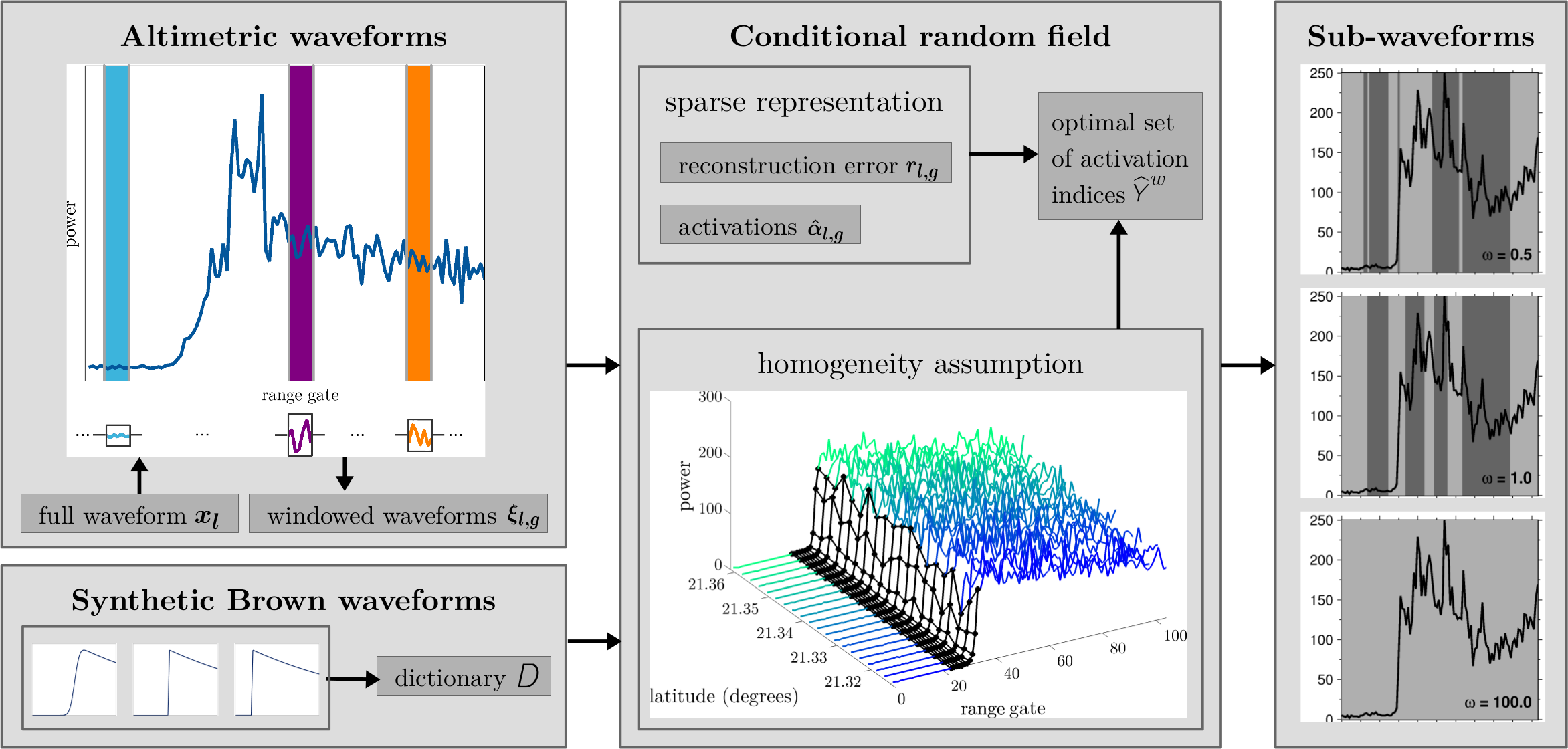}
\caption{Detection Framework: The input is given by altimetric waveforms and synthetic Brown waveforms. Each range gate is represented by a windowed waveform, \ie{}, the range gate's center point and neighboring range gates, as illustrated by blue, violet and orange areas. A conditional random field is formulated, which consists of an unary, data-dependent term computed by sparse representation and a binary term enforcing a smooth partitioning of the waveform. In the conditional random field, the graphical model is constructed by connecting temporally adjacent range gates as well as adjacent range gates within one waveform.
A variation of hyperparameters in the conditional random field result in different partitionings of the entire waveform.}
\label{fig:framework}
\end{figure}

The schematic of our sub-waveform detection framework is illustrated in Fig. \ref{fig:framework}.
The input of the framework are windowed waveforms $\d \xi_{l,g}$ and synthetic Brown waveforms.
These Brown waveforms are the basis waveforms collected in a dictionary $\m D$, which is used for a sparse representation-based modelling of signals. 
Given the input, a conditional random field (CRF) is formulated, which consists of an unary, data-dependent term computed by sparse representation, and a binary term enforcing a smooth partitioning of the entire waveform.
A variation of the weighting between these two terms yield various sets of optimal indices $\hat{\m Y}^w = \left[\hat{\d y}^w_{l,g}\right]$, resulting in different partitionings of the entire waveform.
The framework is flexible regarding the chosen methods, \eg{} the sparse representation can be replaced by other methods such as correlation or other similarity measures.
In the following, detailed explanations to the framework will be provided.

\subsection{Conditional Random Field}
\label{sec:CRF}
In order to perform sub-waveform detection while integrating information about neighboring range gates, we make use of a conditional random field (CRF). 
The range gates are represented in a graph, where each range gate is connected to spatially adjacent range gates within one waveform and to temporally adjacent range gates along the satellite's ground track (see Fig.~\ref{fig:framework}). 
The basic idea is the assignment of each range gate, represented as windowed waveform, to the best-fitting model, for which we use sparse representation in this framework.
Neighbored range gates which are assigned to the same model are summarized to one sub-waveform.
Therefore, all range gates in one sub-waveform follow the same underlying model.
In our approach, the optimal sub-waveform partitioning minimizes the energy functional

\begin{equation}
	\ical E(\m Y) = \sum_{l,g}\ical U(\d \xi_{l,g}, \d y_{l,g}) - w \sum_{l,g,q\in \mathcal Q}\ical B(\d \xi_{l,g}, \d \xi_{l,q}, \d y_{l,g}, \d y_{l,q}),
	\label{eq:CRF}
\end{equation}
where the unary term $\ical U$, depending on the windowed waveforms $\d \xi_{l,g}$, describes the agreement between the measured windowed waveform and a sparse representation model represented by $\d y_{l,g}$.
These models are identified by so-called non-zero activation indices $\m Y = \left[\d y_{l,g}\right]$, indicating which synthetic basis waveforms are used for signal reconstruction within each specific model.
The binary term $\ical B$ depends on both the non-zero activation indices $\d y_{l,g}$ and $\d y_{l,q}$, as well as the windowed waveforms with $q \in \mathcal Q$ indicating the set of direct neighbors of each range gate. The weight between both terms is a hyperparameter and is denoted by $w$.
In our approach, we assume a sub-waveform to be a set of neighboring range gates which are modeled with the same non-zero activation indices, as explained in more detail in the next paragraphs.


\subsubsection{Unary Term - Sparse Representation}
\label{sec:SR}
Generally, in terms of sparse coding (\eg{}, \cite{Olshausen1997}) a $(G \times 1)$-dimensional waveform $\d x_l$ can be represented by a linear combination of a few basis waveforms, which are collected in a $\left(G \times V\right)$-dimensional dictionary $\m D=\left[\d d_v\right]$, $v=1,\ldots, V$, such that $\d x_l = \m D \d \alpha_l + \d \epsilon$ with $ \Vert \d \epsilon \Vert$ being the reconstruction error.
Solving this problem means finding the activation vector $\d \alpha_l$ containing the optimal coefficients, whereas most of the elements are zero.

\begin{figure}[ht]
\centering
	\includegraphics[width=0.6\columnwidth]{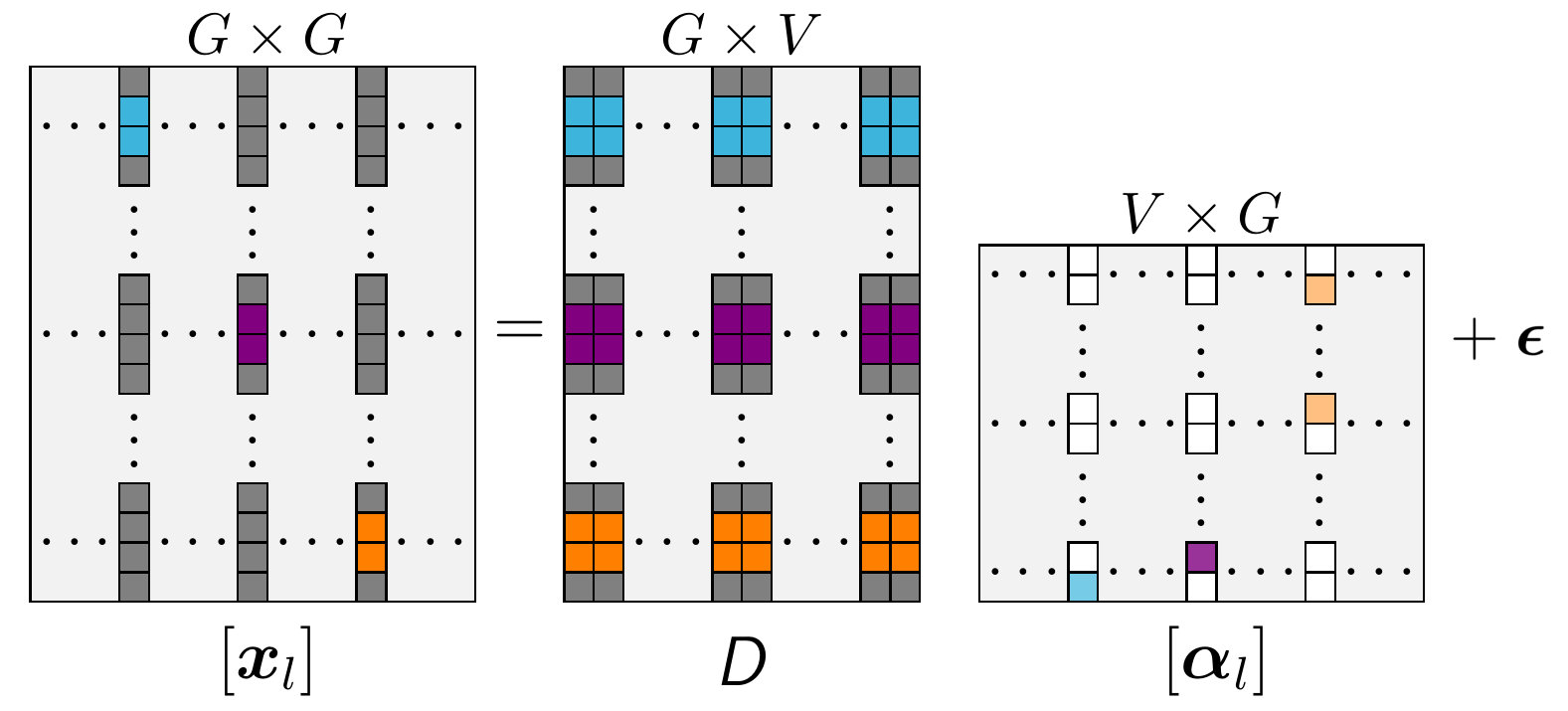}
\caption{Schematic illustration of sparse representation of windowed waveforms. Colors in $\d x_l$ indicate different windowed waveforms $\d \xi_{l,g}$, which are independently sparsely represented. Colors in the dictionary are indicating $\m D_g$, \ie{} the respective rows used for reconstructing the windowed waveform. For sub-waveform detection, neighbored range gates which are represented with the same dictionary elements are grouped to one sub-waveform.}
\label{fig:SR}
\end{figure}

During sub-waveform detection, instead of representing the whole waveform signal, only windowed waveforms will be sparsely represented by windowed basis waveforms.
In this case, each range gate, represented by a windowed waveform, can be assigned to a specific sparse linear combination of windowed basis waveforms constituting the best approximation for it.
In more detail, a windowed waveform $\d \xi_{l,g}$ is sparsely represented using activations $\d \alpha_{l,g}$ and a sub-dictionary $\m D_g$ that contains certain rows of the underlying dictionary $\m D$ (see identical colored entities in Fig. \ref{fig:framework} and Fig. \ref{fig:SR}).
Since the dictionary comprises synthetic Brown waveforms, sub-dictionaries contain specific parts of the synthetic waveforms. 
The indices of non-zero elements in $\d \alpha_{l,g}$, \ie{}, the dictionary elements participating in the reconstruction, are the non-zero activation indices $\d y_{l,g}$ used for defining the models (cmp. Sec. \ref{sec:CRF}). 
It is important to note that $\m D$ is fixed for $L$ consecutive waveforms, such that sub-dictionaries used for temporally neighboring windowed waveforms are identical.

The optimal $\hat {\d \alpha}_{l,g}$ can be formulated as

\begin{equation}
\label{eq:SR}
	\widehat {\d \alpha}_{l,g} = \operatorname{argmin} \Vert \m D_{g} \d \alpha_{l,g} - \d \xi_{l,g} \Vert_2 \qquad \text{subject to} \quad \Vert \d \alpha_{l,g} \Vert_{0} \leq M,
\end{equation}	
where the indices of $M$ non-zero elements in $\widehat {\d \alpha}_{l,g}$ are given by the non-zero activation indices $\d y_{l,g}$.
For a set of optimal activations $\widehat {\d \alpha}_{l,g}$, the reconstruction error of windowed waveform $\d \xi_{l,g}$ will be

\begin{equation}
	r_{l,g} = \Vert \m D_g \widehat {\d \alpha}_{l,g} -\d \xi_{l,g}\Vert_2,
\end{equation}
and $\d r_{l,g} = \left[r_{l,g}\right]$ is the vector collecting the reconstruction errors for all possible sets of dictionary elements.
This optimization can be solved with orthogonal matching pursuit (OMP, \cite{Tropp2006}), which falls in the class of greedy algorithms.
For this, the first dictionary element is chosen to be the one that maximizes the absolute value of the inner product between the dictionary element itself and the sample which is meant to be reconstructed; i.e. we maximize the collinearity. Each further dictionary element is chosen in the same way, however, using the current residual $\d \epsilon$ instead of the sample until the number of the used dictionary elements exceeds $M$. 
The dictionary elements and samples are normalized.
An alternative would be an exhaustive search through all combinations of non-zero activation indices, which however, would be computationally challenging.

In more detail, the unary term in the energy functional penalizes the reconstruction error for a given set of non-zero activation indices and the difference of its activations' sum to $1$, both describing the agreement between data and a specific sparse representation model:

\begin{equation}
	\ical U(\d \xi_{l,g}, \d y_{l,g}) =  \frac{1}{Z_1}\d r_{l,g} + \frac{1}{Z_2} \operatorname{abs}\left(1-\sum_v \widehat{\d \alpha}_{v,l,g}\right),
\end{equation}
where the first term is the normalized reconstruction error obtained by the non-zero activation indices $\d y_{l,g}$, and the second term is the normalized difference to $1$ of the estimated activations $\widehat {\d \alpha}_{l,g}$.
Both unary terms are normalized by the standard deviation $\zeta$ of the values for each range gate, given by $Z_1=\zeta(\d r_{l,g})$ and $Z_2 = \zeta(\operatorname{abs}\left(1-\sum_v \widehat{\d \alpha}_{v,l,g}\right))$, in order to ensure an equal treatment of all range gates.
The sum-to-one penalization serves as regularization to constrain the solution space to reasonable results, which can alternatively be directly incorporated in (\ref{eq:SR}) in a more restrictive way by introducing an additional sum-to-one constraint, as often used in remote sensing \citep{Bioucas-Dias2012}.

A dictionary should have the following properties for our purpose:
First, the elements should have a high approximation ability, and second the choice of used dictionary elements for reconstruction should be unique and stable.
In order to build a suitable dictionary, a set of synthetically generated waveforms is sampled and relevant waveforms are selected to serve as dictionary elements.
We choose the most relevant synthetic waveforms by selecting these ones which are most similar to the estimated waveforms and most dissimilar to each other.
For simulating dictionary elements, the parameters of the Brown model \citep{Brown1977}, such as the waveform amplitude or the epoch, are sampled randomly, where the probability density functions are chosen to resemble empirical distributions of Brown model parameters from a large set of open-ocean echoes from the Envisat or Jason-1/-2 missions.
In this way, the proposed algorithm will be as independent as possible of the chosen altimetry mission.

\subsubsection{Binary Term}
The binary term in \eqref{eq:CRF} serves to incorporate prior knowledge about the spatial relations between adjacent range gates within a single waveform and between temporally consecutive waveforms within one cycle. 
As mentioned before, we assume a sub-waveform to be a set of neighboring range gates which are sparsely represented by a common set of dictionary elements, \ie{} they share the same non-zero activation indices $\d y_{l,g}$, which is set to be the objective in (\ref{eq:CRF}).
Therefore, we prefer neighboring range gates with similar characteristics to be reconstructed with a common set of dictionary elements.
The binary term in \eqref{eq:CRF} is given by

\begin{equation}
	 \ical B(\d \xi_{l,g}, \d \xi_{l,q}, \d y_{l,g}, \d y_{l,q}) = \left\{\begin{array}{cl} \cos\left(\d \xi_{l,g}, \d \xi_{l,q}\right), & \mbox{if\quad} \d y_{l,g} = \d y_{l,q} \\ 0, & \mbox{if\quad} \d y_{l,g}\neq \d y_{l,q} \end{array}\right.. 
\label{eq:CRFg}
\end{equation}
The similarity measure $\cos\left(\d \xi_{l,g}, \d \xi_{l,q}\right)$ relaxes the constraint of the representation of neighboring range gates, \eg{}, in order to consider possible adaptions of the range window by the satellite on-board tracker.

\subsubsection{Sparse Representation-based Conditional Random Field}
As mentioned earlier, the evaluation of all possible sets of non-zero activation indices is computationally difficult. 
However, this information is needed in the CRF to find the best set of indices for all range gates.
To overcome this problem, we optimize the CRF in a greedy manner.
In detail, the search for the optimal estimation of non-zero activation indices in iteration $j=1$ for a windowed waveform is performed over $\mathcal D^{j=1}_{l,g}=\{1,\ldots,V\}$ and fixed after the CRF application. Stopping after a single iteration is identical to the usage of a correlation similarity as unary term. 
For each further iteration $j>1$, the optimal sets of non-zero activation indices will be derived from $\mathcal D^j_{l,g}=\{\{\d y^{j-1}_{l,g}, 1\},\ldots, \{\d y^{j-1}_{l,g}, V\}\}$ with $\d y^{j-1}_{l,g}$ being the optimal set of non-zero activation indices from the previous iteration.
The final CRF that is employed to find an optimal estimation of non-zero activation indices in each iteration is given by the minimizer of the energy

\begin{equation}
	\ical E(\m Y) =  \sum_{l,g} \left( \frac{1}{Z_1}\d r_{l,g} + \frac{1}{Z_2} \operatorname{abs}\left(1-\sum_v \widehat{\d \alpha}_{v,l,g}\right)\right) -
			   w \sum_{l, g, q \in \mathcal Q}\ical B(\d \xi_{l,g}, \d \xi_{l,q}, \d y_{l,g}, \d y_{l,q})\,,
	\label{eq:energy}
\end{equation}
where all entities correspond to the current iteration (index $j$ is omitted for simplicity).
The final optimal set of non-zero activation indices given a specific weight $w$ is denoted by $\widehat{\m Y}^w$.

\section{Sea Surface Height Estimation}
\label{sec:sshe}
In this study, we consider only the effect of the hyperparameter $w$, i.e. the choice of the relative weight between the unary and binary terms in \eqref{eq:CRF}. As it turned out in our numerical experiments, no single choice of $w$ provides optimal results for different cycles, different study sites or altimetry missions. Therefore, we define a range of possible sub-waveforms by varying $w$.
A variation of the hyperparameter $w$ in \eqref{eq:CRF} leads to a variation in the number of detected sub-waveforms. Therefore, we choose a set of reasonable $\mathcal W = \{w_1, \ldots, w_P\}$ hyperparameters. In other words, we derive multiple sub-waveform partitionings, ranging from a very coarse partitioning which includes only a few sub-waveforms to a very fine one which also captures small peaks in separate sub-waveforms. 
For SSH estimation, we use all sub-waveforms which are sufficiently large enough and which can be used for retracking. 
We end up with several equally likely SSHs at each measurement location which could be thought to form a 'point cloud'.
Finally, the Dijkstra algorithm is employed to choose the smoothest combination of these SSHs, such that finally a single SSH at each along-track location is provided.

\subsection{Sea Surface Height}
In the context of this study, we define SSH $\widetilde{h}_{\text{SSH}}$ as given by

\begin{equation}
\widetilde{h}_{\text{SSH}} = a - (R + \Delta_{\text{dry}} + \Delta_{\text{wet}} + \Delta_{\text{iono}} + \Delta_{\text{retr}}),
\label{eq:SSH}
\end{equation}
where $a$ is the satellite's altitude and $R$ is the tracker range related to the fixed tracking gate and provided in the altimeter data records. The atmospheric model corrections $\Delta_{\text{dry}}$, $\Delta_{\text{wet}}$ and $\Delta_{\text{iono}}$, extracted from the SGDR data, refer to the influence of the dry and wet part of the troposphere, as well as the ionospheric influence on the signal. The retracking range correction $\Delta_{\text{retr}}$ is derived from the retracking procedure, described below, by converting the estimated epoch $t_0$ from two-way travel time to range using $\Delta_{\text{retr}}=0.5c t_0$, where $c$ is the speed of light in vacuum.

Additional tidal corrections are applied after the final heights have been selected, to reduce the impact of noise in the tidal corrections on the final height detection. In particular, ocean tide correction will introduce a large noise component in coastal areas which will corrupt the selection of final SSHs through Dijkstra's algorithm. 

Furthermore, we validate our retracking results against tide gauge data with at least hourly temporal resolution without any tidal and barometric corrections applied to remove possible effects from the comparison. For this purpose we add relevant corrections

\begin{equation}
h_{\text{SSH}} =  \widetilde{h}_{\text{SSH}} - \Delta_{\text{ssb}} - \Delta_{\text{set}} - \Delta_{\text{lt}} - 0.468\Delta_{\text{pt}},
\label{eq:SSH2}
\end{equation}
where $\Delta_{ssb}$ is the sea state bias correction which we compute as $5$\% of our retracked SWH. Additionally, $\Delta_{\text{set}}$ is the solid earth tide correction, $\Delta_{\text{lt}}$ is the loading tide and $\Delta_{pt}$ is the pole tide correction. The factor $0.468$ only applies the solid earth part of the pole tide correction, ignoring the part resulting from ocean tides \citep{Fenoglio2015}. 
In other words, we do not apply the ocean tide correction or the inverse barometric correction to render $h_{\text{SSH}}$ being directly comparable to the high rate tide gauge data.


\subsection{Ocean model for Retracking}
\label{sec:retrModel}
For retracking we employ a weighted 3-parameter ocean model \citep{Halimi2013} which is given by

\begin{equation}
\begin{aligned}
x_l(t) = & \frac{A}{2} \left[ 1 + \mbox{erf}\left( \frac{t-t_0 -\psi \sigma^2}{\sqrt{2}\sigma} \right) \right] \\  
& \mbox{exp}\left[-\psi \left( t-t_0 - \frac{\psi \sigma^2}{2}\right) \right] + N_T,
\end{aligned}
\label{eq:retrModel}
\end{equation}
for each two-way travel time $t$, centered on the tracking gate. Here, $N_T$ is the thermal noise which is pre-computed from the first few range gates. The three fitted model parameters represent the amplitude $A$, which is related to backscatter, the rise time of the leading edge $\sigma$ which can be converted to SWH, as well as the epoch or retracking gate $t_0$ which refers to the position of the mid point of the leading edge. Additionally, $\psi$ is defined by

\begin{equation}
\begin{aligned}
\psi &= \frac{4 c}{\gamma h} \text{ } \frac{1}{1+a/R_e} \cos(2\phi),\\
\gamma &= \frac{\sin^2(\theta)}{\ln 4},
\end{aligned}
\label{eq:retrModelAlpha}
\end{equation}
where $c$ is the speed of light in vacuum, $R_e$ is the radius of the Earth \citep{Deng2003} and $\phi$ is the off-nadir pointing angle which is assumed to be zero in our application. The antenna beamwidth parameter $\gamma$ is defined in \citep[Eq. (4)]{Brown1977} and can be computed from the beamwidth $\theta$ of the altimeter instrument.

\subsection{Shortest-Path Algorithm for Finding the Best Set of Sea Surface Heights}
\label{subsec:dijk}
\begin{figure}[ht]
\centering
	\includegraphics[width = 0.5\columnwidth]{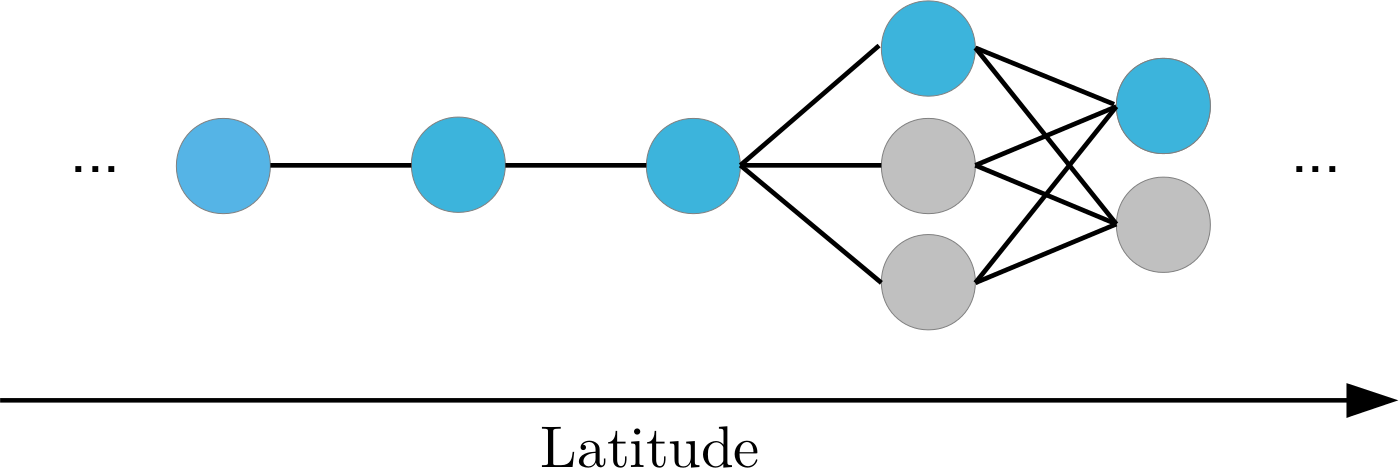}
	\caption{Schematic illustration of optimal sea surface height estimation employing Dijkstra algorithm. Dijkstra finds the optimal path (illustrated in blue) through equally likely successively arranged sea surface heights. where edge weights are derived from the difference between heights of 2 nodes.}
	\label{fig:dijkstra}
\end{figure}

\begin{figure}[ht]
\centering
	\includegraphics[width = 0.8\columnwidth]{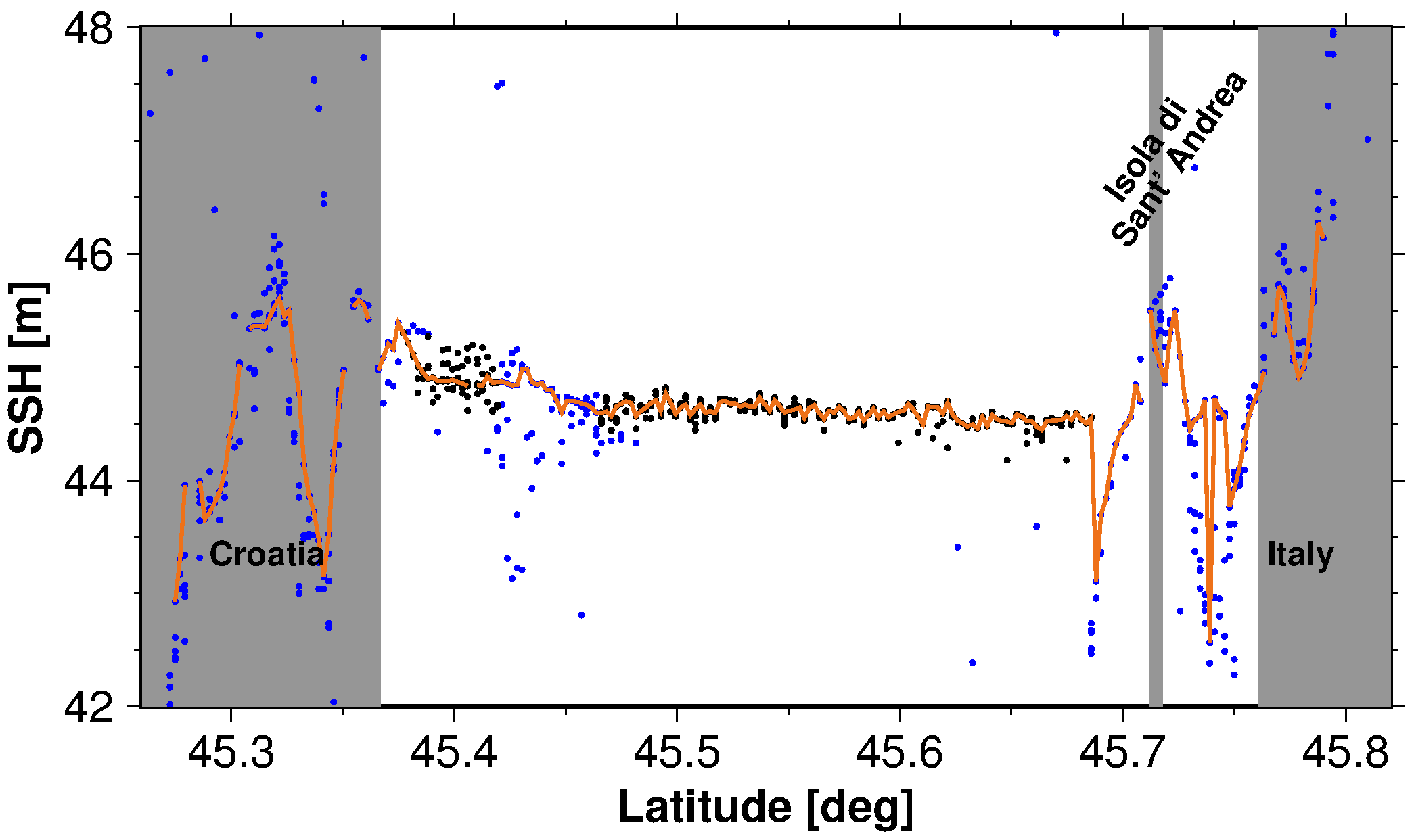}
	\caption{Exemplary SSH pointcloud from the Triest study site as part of our proposed STAR algorithm. The black points are selected by the RANSAC algorithm to estimate the linear model in order to discard outliers. Afterwards all points within range of the linear model are used in the Dijkstra algorithm to find the best (smoothest) set of SSHs (orange line).}
	\label{fig:pcmeth}
\end{figure}

The problem that we discuss here can be viewed as an optimization problem with certain constraints. Given now a  'point-cloud' of (equally) possible SSHs for each groundtrack-point, we need to realize a consistent SSH at each measurement position by finding the optimal SSH candidates. Before applying a shortest-path algorithm, we remove outliers from the 'point cloud' through the RANdom SAmple Consensus (RANSAC) algorithm \citep{Fischler1981} while assuming a linear model of SSH change in along-track direction. Sea level change is not linear and thus, we apply the RANSAC algorithm with a threshold of \unit[3]{m} to a moving window covering about \unit[20]{s} of measured data. By selecting the moving window to cover \unit[20]{s} we make sure that we always include a relatively large portion of water, especially at land-ocean transitions. All points that deviate too far from the linear model estimated by the RANSAC algorithm are discarded and the resulting set of accepted SSHs is then used to find the shortest path and estimating the optimal SSHs at each measurement location as illustrated in Fig. \ref{fig:dijkstra} and Fig. \ref{fig:pcmeth}. Here, we chose Dijkstra algorithm \citep{Dijkstra1959}, but other shortest-path algorithms would also be possible. Dijkstra's method requires one to choose edge weights between individual connected nodes. In our application, we chose the height differences between connected nodes as edge weights, thus, favoring smaller height changes over larger ones.
For the start and end point of the Dijkstra graph we use the first and last sea surface height at the start and end position. 

\section{Results}
\label{sec:experiments}

\subsection{Setting}
We compare the STAR algorithm to existing retracking algorithms, such as the standard range derived from a MLE4 retracking method provided in the SGDR data, the $30$\%-threshold retracker (\thresh{}, \cite{Martin1983}) and the equally weighted 3-parameter ocean model (\wtp{}, see Eq. \eqref{eq:retrModel}). Furthermore, the specialized coastal algorithms, such as the Adaptive Leading Edge Subwaveform retracker (\ales{}, \cite{Passaro2014}) and the Improved Threshold Retracker (\itr{}, \citep{Hwang2006}), the latter combined with a threshold of $50$\%, are considered. 

When implementing the STAR method as described above, we chose to set the neighborhood of the windowed waveforms $\d \xi$ to $N_{\xi} = 5$. For the number of non-zero elements in \eqref{eq:SR} we select $M=2$. Values of $M > 2$ lead to significantly increased computation times while a larger combination of the limited number of dictionary elements might result in less clearly defined sub-waveforms from unique combinations of the basis elements. To avoid finding an optimal weighting parameter $w$ for each measurement region, we run \eqref{eq:CRF} for five different choices of $w \in \mathcal W = \{0.1, 0.5, 1, 2, 100\}$.
This will result in five partitionings of the total waveform into sub-waveforms, ranging from a very fine partitioning to a very coarse one. 
The parameters for the computation of our basis elements are generated based on the average estimated parameters from an application of the $50$\%-threshold retracker to the current block of waveforms (20 waveforms per block in our framework). 
The mean epoch and amplitude serve as input into the basis element generation to produce $1000$ waveforms by randomly varying the mean parameters: the amplitude is randomly varied by about $10$\% of its mean threshold-retracked value. 
The epoch is randomly sampled using a Gaussian weighted distribution with its maximum at the mean threshold-retracked value. 
The waveheight is randomly sampled from the range \unit[0]{m} to \unit[12]{m}, which is a most common range of waves. 
The cross-correlation coefficients between the obtained $1000$ generated waveforms are computed and only the 15 waveforms which are most distinct from each other are kept and form the dictionary elements.
In the following we utilize all the detected sub-waveforms to derive the results for \STAR{}.    

Our \STAR{} algorithm, \thresh{}, \wtp{} and \itr{} have been implemented in our in-house C++ altimetry toolbox. As \itr{} provides heights for each detected leading edge, we assumed the first detected leading edge to yield the correct retracking parameters, as we do not have any prior information on the heights. The \sgdr{} SSHs are extracted from the GDR data and \ales{} retracked ranges are extracted from external GDR data. The SSHs from all retracking methods have been processed in the same way.

\subsection{Sub-waveform Detection}

\begin{figure}[ht]
\centering
	\includegraphics[width = 0.98\columnwidth]{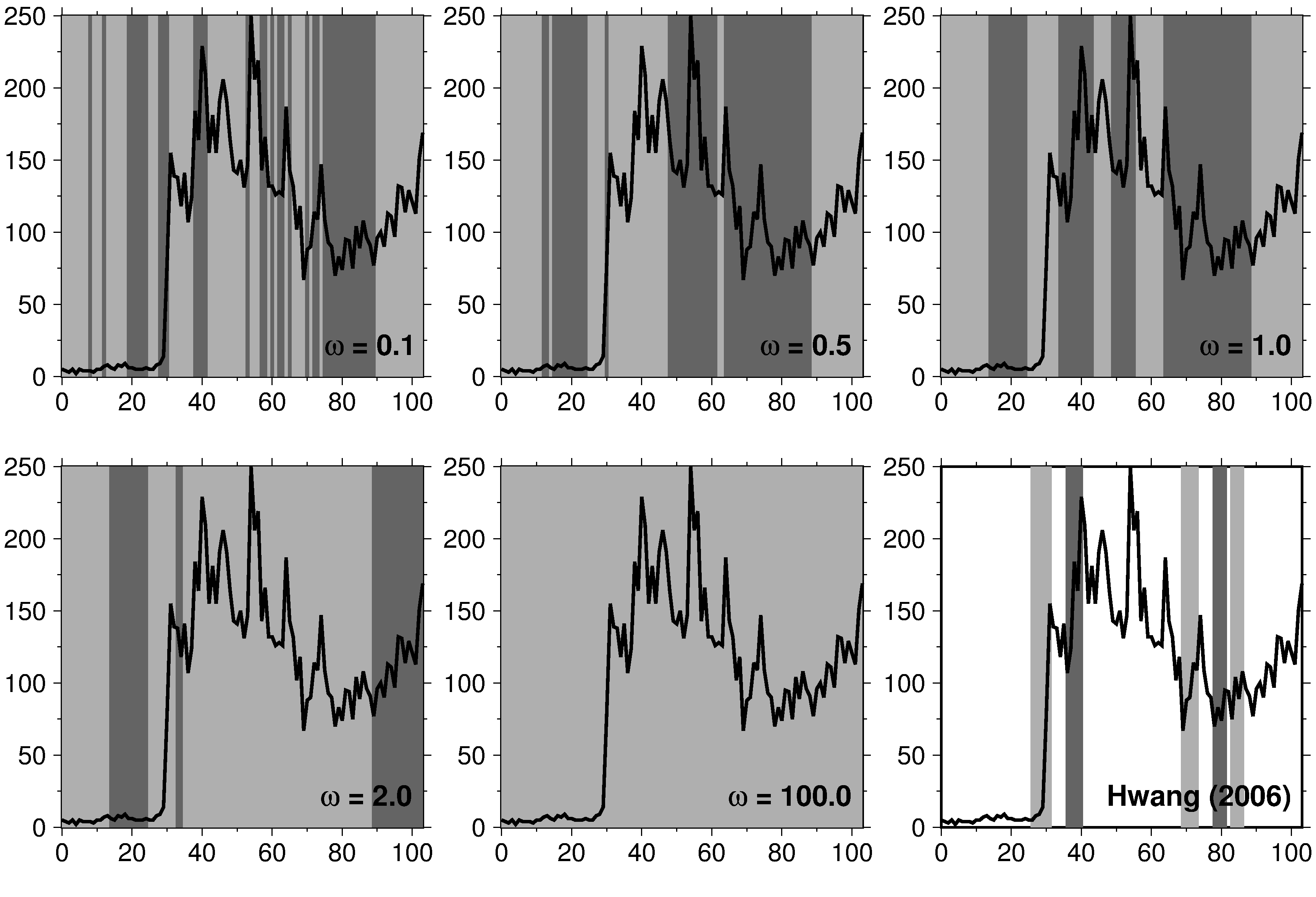}
	\caption{Sub-waveform partitioning using our method for weights from the set $\mathcal W = \{0.1, 0.5, 1, 2, 100\}$ inserted into  \eqref{eq:CRF}. The waveform is an arbitrarily selected coastal waveform from Jason-2 pass 196, cycle 69. Grey shaded background depicts the individual sub-waveforms while a white background indicates no sub-waveform in that part of the total waveform.}
	\label{fig:swf1}
\end{figure}

First, we compare the sub-waveforms detected by STAR resulting from five different weights $w \in \mathcal W$ in \eqref{eq:CRF} to sub-waveforms from the method by \citet{Hwang2006} for an arbitrarily chosen waveform measured off the coast of Croatia (Fig. \ref{fig:swf1}). 
The waveform parts that have not been detected as part of a leading edge by the method presented by \citet{Hwang2006} are indicated by the white background. In contrast, individual sub-waveforms are depicted in alternating shades of grey. For this waveform, five potential leading edges have been identified by the method of \cite{Hwang2006}, each of which can be utilized to derive potential sea surface heights. 
When comparing the partitioning of the total waveform for the different weights $w \in \mathcal W$ by \STAR{}, lower weights lead to a significantly increased number of identified sub-waveforms. In some cases single sub-waveforms contain only one range gate since the binary term in \eqref{eq:CRF} is weighted significantly lower than the unary term. With increased weight $w$, the similarity constraint is enforced and the sub-waveform size increases. For a weight $w=100$, the identified sub-waveform corresponds to the entire waveform, which leads to including the standard case of retracking the complete waveform in our derived point clouds. 

\begin{figure}[ht]
\centering
	\includegraphics[width = 0.8\columnwidth]{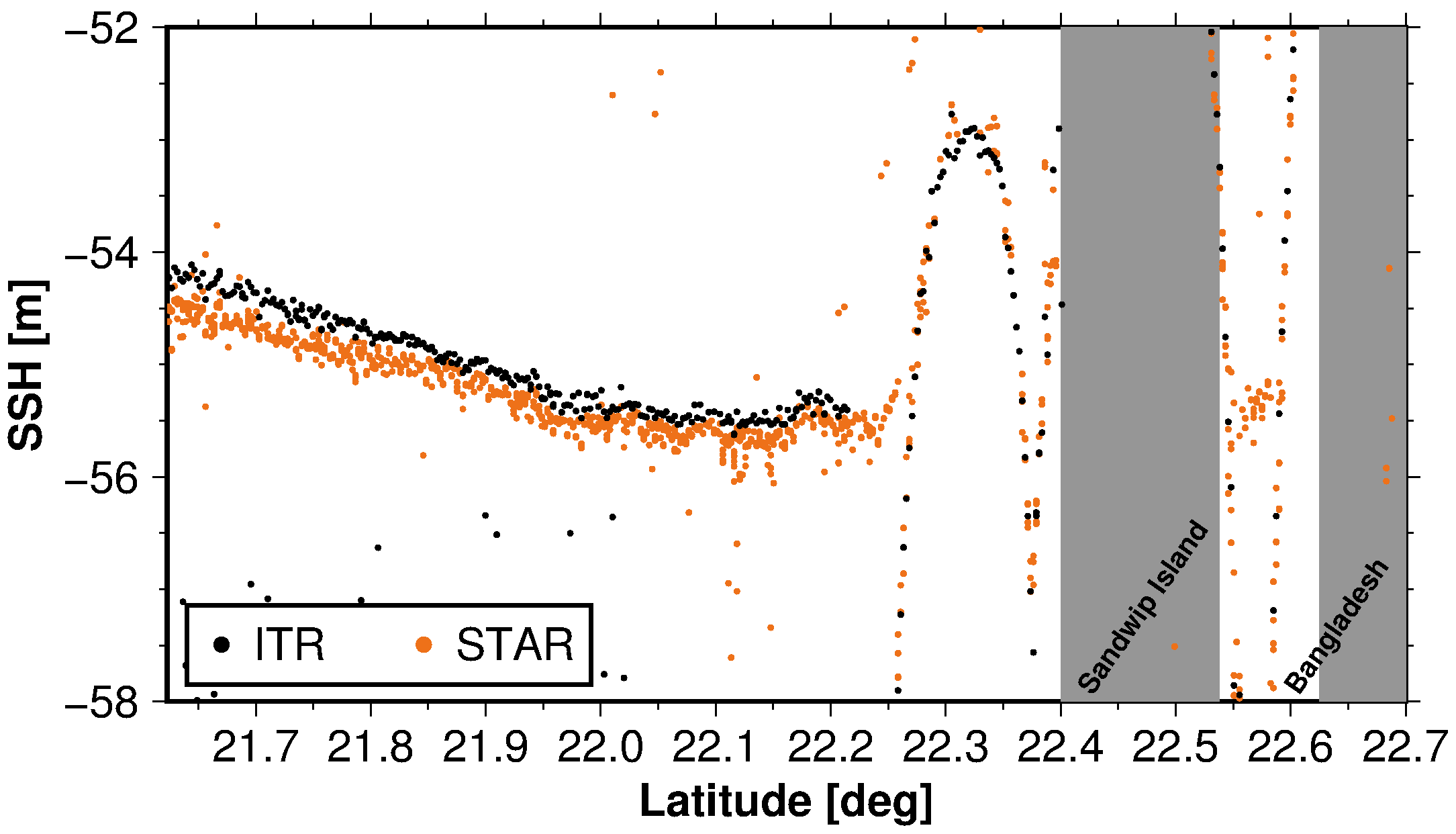}
	\caption{Point cloud of SSHs resulting from \STAR{} compared to \itr{} \citep{Hwang2006}.}
	\label{fig:swf2}
\end{figure}

In a next step, we compare the resulting point clouds of \STAR{} (orange points, Fig. \ref{fig:swf2}) to the point cloud derived using \citet{Hwang2006} algorithm (black points, Fig. \ref{fig:swf2}) for one exemplary cycle during low tide conditions at the Bangladesh site. 
A small bias over the open ocean can be identified between both methods, which is related to the chosen threshold of $50\%$ used for \itr{}. For the sandbank near \unit[22.325]{$^\circ$N} both point clouds agree well. However, prior to reaching the sandbank the points derived from \itr{} drop rapidly by about \unit[10]{m} to a level of approximately \unit[-64]{m} (outside the plot boundaries), while the point cloud based on \STAR{} becomes less dense but still preserves enough meaningful SSHs of about \unit[-55.5]{m} up to the beginning of the sandbank. 
Between Sandwip Island and Bangladesh mainland, \STAR{} is able to derive SSHs from the detected sub-waveforms, while \itr{} is more influenced by land returns disturbing the retrieved waveform, which results in no meaningful SSHs that can be detected during the low tide conditions in this case.

\subsection{Retracked SSHs}
\label{subsec:retrSSH}
\begin{figure*}[ht]
\centering
	\includegraphics[width = 0.95\textwidth]{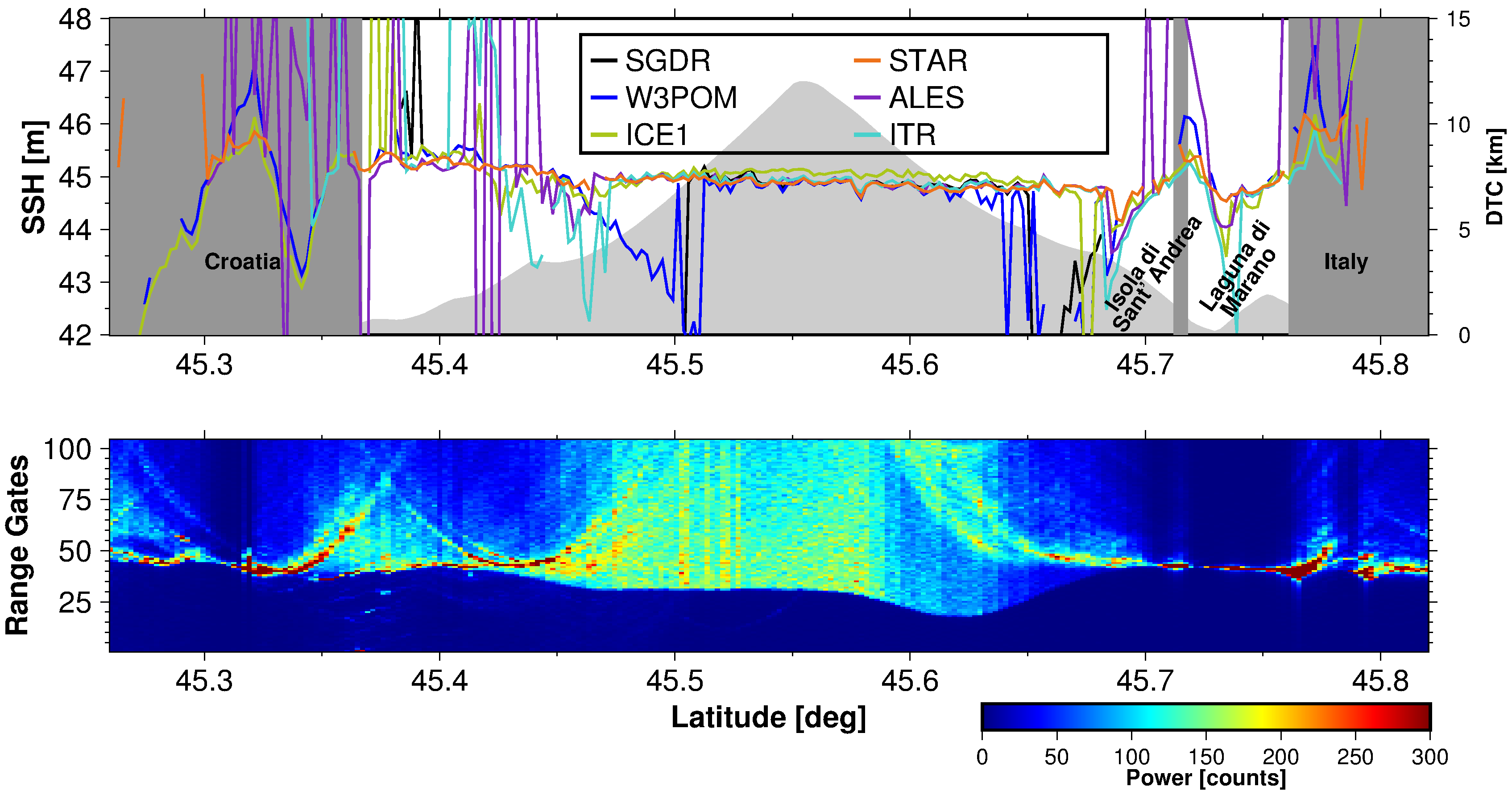}
	\caption{(top) Comparison of SSH derived by various retrackers along the Jason-2 pass 196 in the Gulf of Trieste for an arbitrarily chosen cycle (cycle 165). The distance to the nearest coastline (DTC) is provided in light gray. (bottom) Radargram of the corresponding waveforms of inside the study area. The waveform return power is given color coded and has been capped at a power of 300 for visual reasons.}
	\label{fig:resTssh}
\end{figure*}

For validation of the retracked SSHs, we compare STAR SSHs to retracked SSHs from various conventional and coastal methods.
The top of Fig. \ref{fig:resTssh} shows retracked SSHs derived for one arbitrarily selected cycle~165 at almost high tide at the Trieste study site. 
The bottom part of Fig. \ref{fig:resTssh} shows the corresponding return waveforms at each measurement location. 

Over the open ocean between $\sim$\unit[45.46]{$^\circ$N} and \unit[45.58]{$^\circ$N}, we find the waveforms corresponding well to the theoretical Brown model and the SSHs from all retracking algorithms agree well. 
However, \thresh{} shows a small bias compared to the other retracking algorithms which is likely due to the $30$\%-threshold.
At about \unit[5-7]{km} off the nearest coast in the northern and southern part of the study site, the SSHs based on the \wtp{} and \sgdr{} indicate a rapid drop in sea level. 
This drop is related to the influence from peaks, resulting from land influence from the Croatian and Italian mainlands. 
The coastal methods \STAR{}, \ales{} and \itr{} are not influenced by the peaks as they consider only parts of the total waveform that do not include the peaks at these measurement positions.

In the southern coastal area (\unit[45.37]{$^\circ$N} to \unit[45.46]{$^\circ$N}), the SSHs from \wtp{}, again, generally agree with the SSHs from \thresh{} and \STAR{} with the latter showing less noisy along-track variations and providing SSHs right up to the coast of the Croatian mainland (Fig. \ref{fig:resTssh}). Between \unit[2 to 4]{km} off the Croatian coast the SSHs based on \ales{} and \itr{} show significant outliers which are related to the peak, due to the land influence of the Croatian mainland, being located very close to the leading edge resulting in biased estimations. Additionally, smaller peaks preceding the leading edge can be detected (Fig. \ref{fig:resTssh}, bottom) which are identified as potential leading edges by the \itr{} algorithm and, thus, result in outliers due to our assumption to utilize the first identified leading edge in the \itr{} algorithm. Similar behavior can be observed from the \thresh{} algorithm, e.g. near $\sim$\unit[43.75]{$^\circ$N}, where the peaks preceding the leading edge lead to significant outliers.

In the northern coastal area (\unit[45.64]{$^\circ$N} to \unit[45.71]{$^\circ$N}), we find good agreement between the SSHs based on \ales{}, \itr{} and \STAR{}, with a few significant outliers from \ales{} over the last two kilometers off the nearest coast. At about \unit[3.5]{km} off the Isola di Sant' Andrea, a drop in SSH can be detected from all three methods with \itr{} showing the strongest drop by several meters while the SSHs from \STAR{} only drop less than \unit[1]{m} (Fig. \ref{fig:resTssh}) due to relatively broad peaks located directly at the leading edge.

Over the relatively shallow Laguna di Marano all methods, except for \sgdr{}, provide a sea level similar to the open ocean level during high tide, despite of the waveforms consisting of strong specular peak shapes (Fig. \ref{fig:resTssh}, bottom).

\subsection{Repeatability of the Retracked SSHs}
\label{sec:repeat}
\begin{figure*}[ht]
\centering
	\includegraphics[width = 0.95\textwidth]{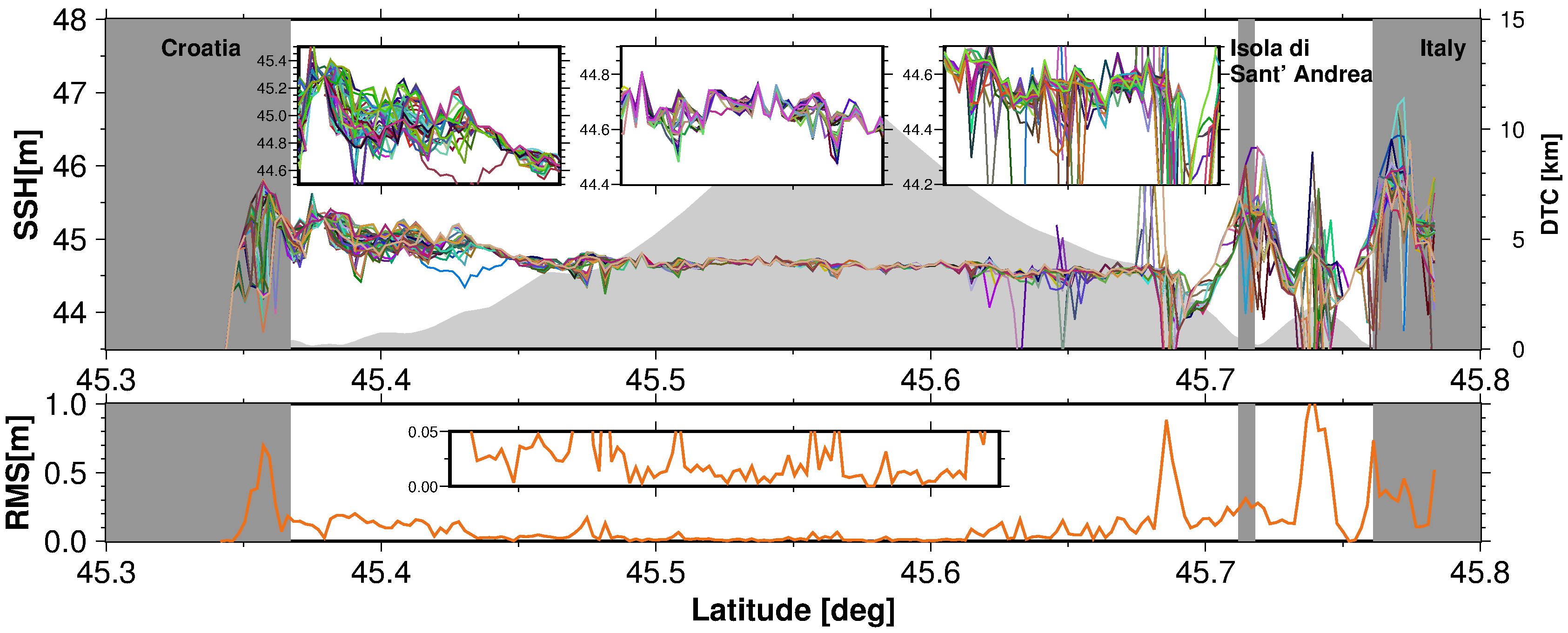}
	\caption{Repeatability of \STAR{}. (top) SSHs obtained from 1000~runs on cycle 69, pass 196 at our Trieste study site. Additionally, we show zoomed in sub-plots for coastal and open ocean regions. Colors are chosen randomly. (bottom) Corresponding root mean square difference (RMS) derived from 1000~runs where open ocean region is also provided in a zoomed in sub-plot.}
	\label{fig:repeat}
\end{figure*}

The \STAR{} method utilizes a randomized dictionary for sub-waveform decomposition; therefore it makes sense to investigate whether this leads to an uncertainty in the final SSHs. To answer this question we conduct a Monte Carlo study: we ran the \STAR{} algorithm 1000~times, utilizing the arbitrarily selected cycle 69 of pass 196 of the Jason-2 mission. Resulting SSHs are shown in Fig. \ref{fig:repeat} (top) with additional zoom-in regions for two coastal areas, as well as for the open ocean. The bottom part of the figure displays the corresponding root mean square difference (RMS) ranging from about \unit[0 to 1]{m}, where the open ocean area is also given in an additional sub-plot for a range of \unit[0 to 5]{cm}.

Over the open ocean area, the variability along the groundtrack is in the range of about \unit[15-20]{cm} (Fig. \ref{fig:repeat}, top). Here, the repeatability of \STAR{} is given by a RMS of less than \unit[5]{cm} (Fig. \ref{fig:repeat}, bottom).

In the southern coastal area, the along-track variability is in the range of \unit[60]{cm} revealing a slight increase in sea level towards the Croatian coast (Fig. \ref{fig:repeat}, top). Due to the land influence on the shape of the waveform as is visible in the parabolas (Fig. \ref{fig:resTssh}, bottom), the retracking results are sensitive with respect to the size of the individual detected sub-waveforms. 
Nonetheless, we still find good repeatability in this region with RMS of less than \unit[20]{cm} (Fig. \ref{fig:repeat}, bottom), which is less compared to the along track variability and significantly less compared to the relatively large outliers produced by other retracking algorithms in this region (see Fig. \ref{fig:resTssh}, top). 

Part of the variability in our current state of the algorithm is due to the Dijkstra algorithm, which is employed to find the best SSHs in our point cloud. Single points at a measurement location can have significant influence on the chosen path since we only allow edge connections between neighboring locations. For example at $\sim$\unit[45.375]{$^\circ$N}, we find a standard deviation of almost zero (Fig. \ref{fig:repeat}, bottom) and in the top plot of Fig. \ref{fig:repeat} it is possible to identify a single point where the SSH-tracks of all runs intersect. 
Since all paths obtained by Dijkstra algorithm include this point, preceding and succeeding SSHs
are influenced and tend to be close to the SSH at this point.

In the northern coastal region, the general along-track variability is about in the same range of \unit[20]{cm} as compared to the open ocean area. The repeatability in the shallow waters off the Isola di Sant' Andrea is similar to the Croatian coast in the south with a RMS of less than \unit[20]{cm}. However, we find some significant outliers in some runs which are related to a weak leading edge and strong peaks close to the leading edge (e.g., Fig. \ref{fig:resTssh}, bottom). Over the Laguna di Marano a large variability can be detected (Fig. \ref{fig:repeat}) due to strong specular peak waveforms. Small changes in the detected sub-waveforms will have a significant impact on the derived SSHs. 

\subsection{Comparison to Tide Gauge Data}
\label{subsec:compTide}
In this experiment we compare the SSH estimated using various retracking methods with tide gauge data for Trieste, Italy and Chittagong, Bangladesh.
For comparison we utilize cycles 1 to 230 of Jason-2 data acquired over the described study sites (see Sec. \ref{sec:data}), since for this period hourly tide gauge data as well as data obtained by \ales{} is available. The tide gauge data between July 2009 and December 2014 are interpolated to the times of crossing for each cycle.
As the tide gauge data is not corrected for tidal or atmospheric pressure effects, we also do not apply these corrections to the altimetry data by employing \eqref{eq:SSH2}.
We then remove outliers from the altimetric SSHs (deviation $>$ \unit[3]{m} from mean SSH) and evaluate correlation and difference RMS with respect to tide gauge data.
Due to the large number of available Jason-2 cycles, we set the minimum number of cycles that are required to derive reliable correlations and RMS to 50.

\begin{figure*}[ht]
\centering
	\includegraphics[width = 0.95\textwidth]{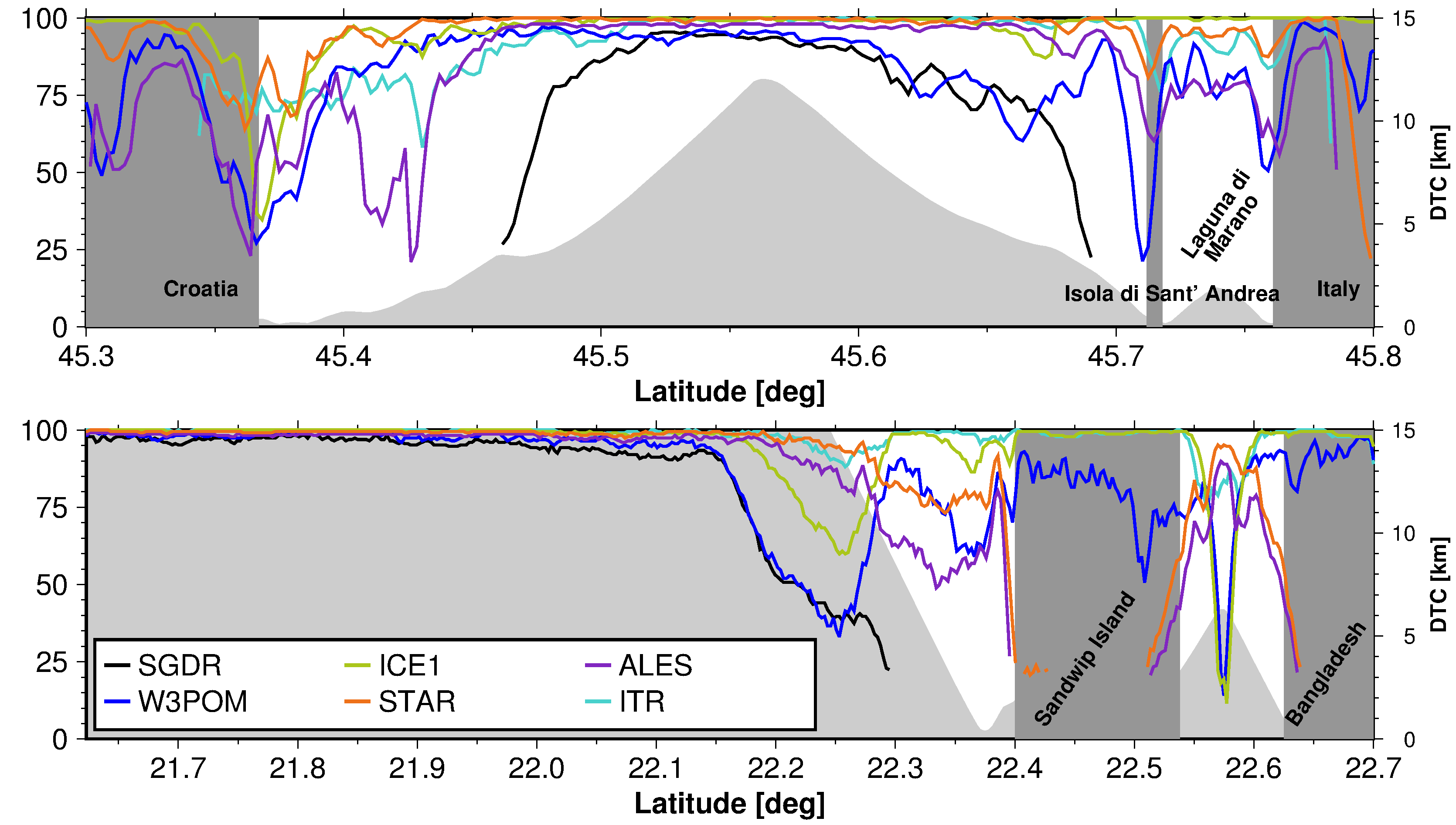}
	\caption{Percentage of the total number of available Jason-2 cycles (227) after applying outlier detection and minimum number of cycles requirement. (top) Study site at the Gulf of Trieste. (bottom) Study site at the coast of Bangladesh. The distance to the nearest coastline (DTC) is provided in light gray.}
	\label{fig:numMC}
\end{figure*}

The percentage of all cycles which meet the criteria above is shown in Fig. \ref{fig:numMC}. The median results of the following sections are summarized in Tables \ref{tab:resultsT} and \ref{tab:resultsB}. 

\begin{table}
\centering
\caption{Median values of the different quality measures for the study site in Trieste. See also top plots in Figs. \ref{fig:korr}, \ref{fig:numTot} and \ref{fig:rms}. Here, $\rho$ represents the correlation, $N$ the percentage of retained cycles and $\sigma$ the RMS.}
\begin{tabular}{lccccccccc}
\hline
\textbf{Retracker} & \multicolumn{3}{c}{Open Ocean} & \multicolumn{3}{c}{Croatian Coast} & \multicolumn{3}{c}{Italian Coast} \\ 
 & $\rho$ [-] & $N$ [\%] & $\sigma$ [m] & $\rho$ [-] & $N$ [\%] & $\sigma$ [m] & $\rho$ [-] & $N$ [\%] & $\sigma$ [m] \\ \hline
SGDR & 0.62 & 85 & 0.43 & - & - & - & 0.27 & 8 & 1.35 \\
W3POM & 0.53 & 85 & 0.53 & 0.54 & 59 & 0.53 & 0.27 & 7 & 1.58 \\
ICE1 & 0.90 & 99 & 0.15 & 0.46 & 42 & 0.74 & 0.63 & 74 & 0.44 \\
ITR & 0.86 & 99 & 0.17 & 0.34 & 36 & 0.83 & 0.69 & 88 & 0.32 \\
ALES & 0.86 & 97 & 0.16 & 0.17 & 3 & 1.62 & 0.26 & 60 & 0.50 \\
STAR & 0.90 & 99 & 0.15 & 0.73 & 81 & 0.28 & 0.80 & 98 & 0.22\\
 \hline

\end{tabular}
\label{tab:resultsT}
\end{table}

\begin{table}
\centering
\caption{Median values of the different quality measures for the study site at the coast of Bangladesh. See also bottom plots in Figs. \ref{fig:korr}, \ref{fig:numTot} and \ref{fig:rms}. Here, $\rho$ represents the correlation, $N$ the percentage of retained cycles and $\sigma$ the RMS.}
\begin{tabular}{lccccccccc}
\hline
\textbf{Retracker} & \multicolumn{3}{c}{Open Ocean} & \multicolumn{3}{c}{Sandbank} & \multicolumn{3}{c}{Channel} \\ 
 & $\rho$ [-] & $N$ [\%] & $\sigma$ [m] & $\rho$ [-] & $N$ [\%] & $\sigma$ [m] & $\rho$ [-] & $N$ [\%] & $\sigma$ [m] \\ \hline
SGDR & 0.94 & 93 & 0.40 & 0.64 & 15 & 1.06 & - & - & - \\
W3POM & 0.92 & 96 & 0.56 & 0.31 & 24 & 1.82 & 0.29 & 28 & 1.33 \\
ICE1 & 0.95 & 99 & 0.43 & 0.50 & 55 & 1.43 & 0.22 & 29 & 1.34 \\
ITR & 0.97 & 99 & 0.37 & 0.63 & 72 & 1.15 & 0.40 & 34 & 1.33 \\
ALES & 0.95 & 98 & 0.38 & 0.14 & 47 & 1.38 & 0.46 & 27 & 1.36 \\
STAR & 0.97 & 99 & 0.33 & 0.88 & 78 & 0.64 & 0.88 & 68 & 0.72\\
 \hline

\end{tabular}
\label{tab:resultsB}
\end{table}

\subsubsection{Overall Correlation}

\begin{figure*}[ht]
\centering
	\includegraphics[width = 0.95\textwidth]{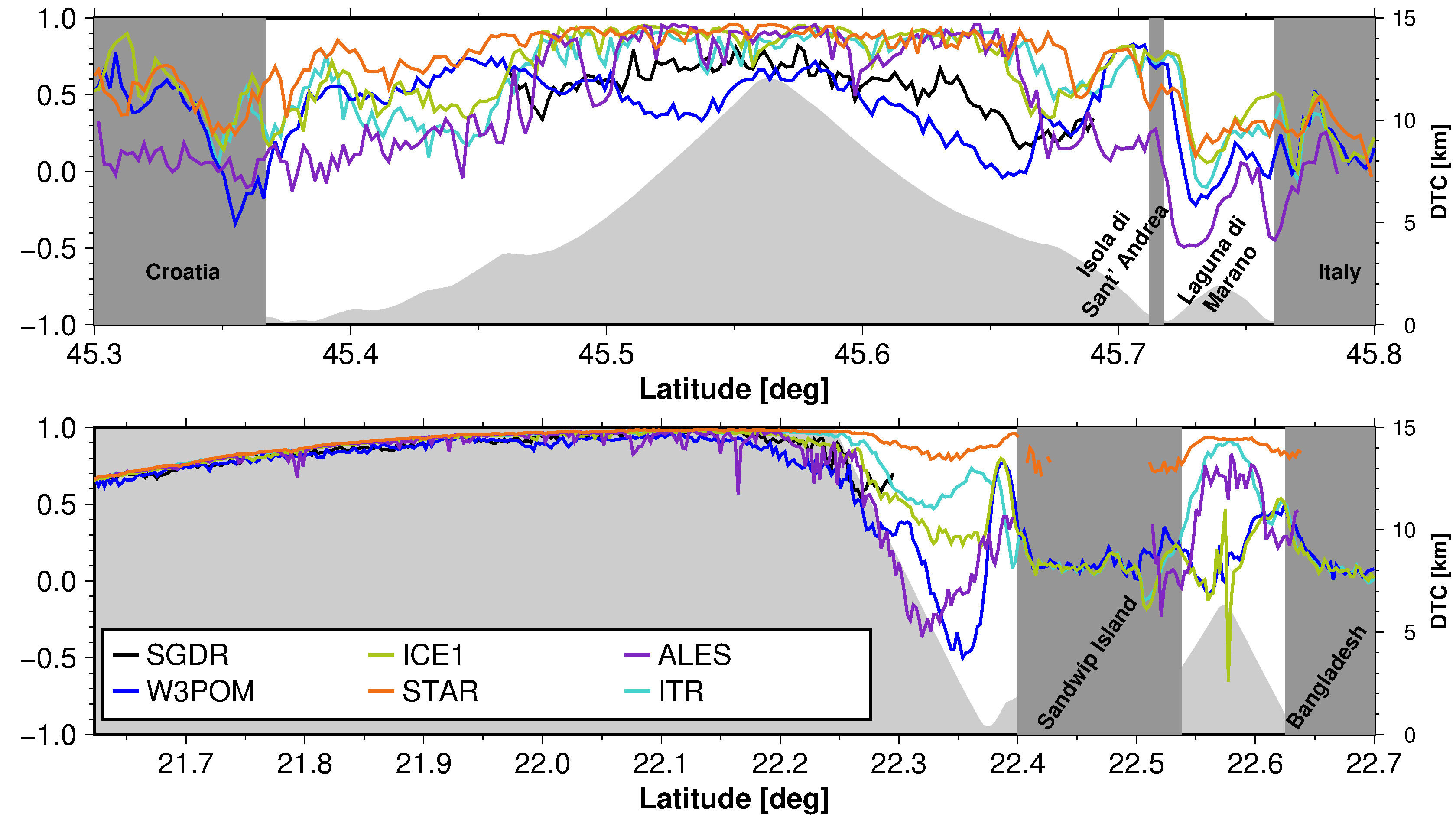}
	\caption{Correlation $\rho$ of SSHs derived from several retracking algorithms, including our STAR method, with hourly tide gauge data. (top) Study site at the Gulf of Trieste. (bottom) Study site at the coast of Bangladesh. The distance to the nearest coastline (DTC) is provided in light gray.}
	\label{fig:korr}
\end{figure*}

We investigate the overall correlation $\rho$ with tide gauge data at each \unit[20]{Hz} along-track position, using all available cycles (Fig. \ref{fig:numMC}). The results for the Jason-2 mission are shown in Fig. \ref{fig:korr} for the study site at the Gulf of Trieste (top) and the coast of Bangladesh (bottom).

Over the open ocean area at the Trieste study site (Fig. \ref{fig:korr}, top), SSHs obtained by \thresh{}, \itr{}, \ales{} and \STAR{} show correlations of more than 0.8 with the tide gauge time series. For \wtp{} and \sgdr{} we find correlations between 0.3 and 0.7. Over the open ocean, southward of \unit[22.2]{$^\circ$N}, in the Bangladesh study area (Fig. \ref{fig:korr}, bottom), the SSHs derived from the considered retracking methods agree well to each other. In the South, the coastal shelf transitions towards the deep ocean and the correlation begins to drop to about $\rho=0.6$ at the border of the study site. Over the deep ocean, the correlation with tide gauge data from the Chittagong station drops rapidly, as already seen by \cite{Kusche2016}. Over the central coastal shelf up to about \unit[15]{km} off the coast, we find correlations of up to $100$\% between retracked SSHs and tide gauge data. In the immediate coastal areas of the Trieste and Bangladesh study sites, the MLE4 algorithm used to derive the standard ranges in the SGDR data did not converge and consequently no \sgdr{} SSHs are available in these regions.

Towards the Croatian coast, we find $\rho_{\text{\thresh{}}}$ and $\rho_{\text{\itr{}}}$ to decline to a level of 0.3 to 0.5 (Fig. \ref{fig:korr}, top) which agrees to $\rho_{\text{\wtp{}}}$ in this region, while $\rho_{\text{\ales{}}}$ shows a rapid decline to a level of 0 to 0.2 at about \unit[3]{km} off the coast. $\rho_{\text{\STAR{}}}$ also declines towards the coast to a level of 0.5 to 0.7. 

At the Italian coast, we again find a decline in overall correlation of the retracked SSHs with the tide gauge time series (Fig. \ref{fig:korr}, top). 
About \unit[4 to 5]{km} off the Isola di Sant' Andrea, $\rho_{\text{\thresh{}}}$ and $\rho_{\text{\ales{}}}$ rapidly decline to a level of 0 to 0.2. 
While $\rho_{\text{\ales{}}}$ remains at this level, $\rho_{\text{\thresh{}}}$ increases again to a level of 0.8 right at the coast, which agrees well to $\rho_{\text{\wtp{}}}$, $\rho_{\text{\itr{}}}$ and $\rho_{\text{\STAR{}}}$. 
Over the Laguna di Marano, we generally find a lower correlation level of 0 to 0.3 or even negative correlation in the southern part of the Laguna for all retracked SSHs. 

Starting about \unit[15]{km} off the coast of Sandwip Island at the Bangladesh study site, the overall correlations for all retrackers decline (Fig. \ref{fig:korr}, bottom). 
Over the sandbank area, $\rho_{\text{\thresh{}}}$ drops to a level of 0.3 to 0.4, $\rho_{\text{\itr{}}}$ shows a more moderate drop to a level of 0.5 to 0.7 and $\rho_{\text{\STAR{}}}$ indicates only a small drop to 0.8, which is achieved due to automatically removing low tide conditions where the sandbank height does not fit to the conditions imposed by the RANSAC algorithm before applying Dijkstra's algorithm (see Sec. \ref{subsec:dijk}). 
Over the strip of about \unit[13]{km} of open water which is part of the estuary of the Ganges-Brahmaputra-Meghna Delta, we find $\rho_{\text{\wtp{}}}$ and $\rho_{\text{\thresh{}}}$ at nearly zero, while for \ales{}, \itr{} and \STAR{}, $\rho_{\text{\ales{}}} \approx 0.7$, $\rho_{\text{\itr{}}} \approx \text{0.8 to 0.9}$ and $\rho_{\text{\STAR{}}} \approx 0.9$ are found in the central part of this region. Lower correlations of \wtp{} and \thresh{} while showing similar number of retained cycles and RMS (Table \ref{tab:resultsB}) are related to the fact that both methods utilize the total waveform which mainly consists of a strong peak only. For \wtp{} this will often lead to divergence of the estimation. The heights from the \thresh{} algorithm will be influenced by the signal shape of the waveform. This leads to varying parts of the waveform being used to derive the amplitudes and consequently to inconsistencies when deriving the range correction.

\subsubsection{Number of Retained cycles}
\label{sec:retained}
\begin{figure*}[ht]
\centering
	\includegraphics[width = 0.95\textwidth]{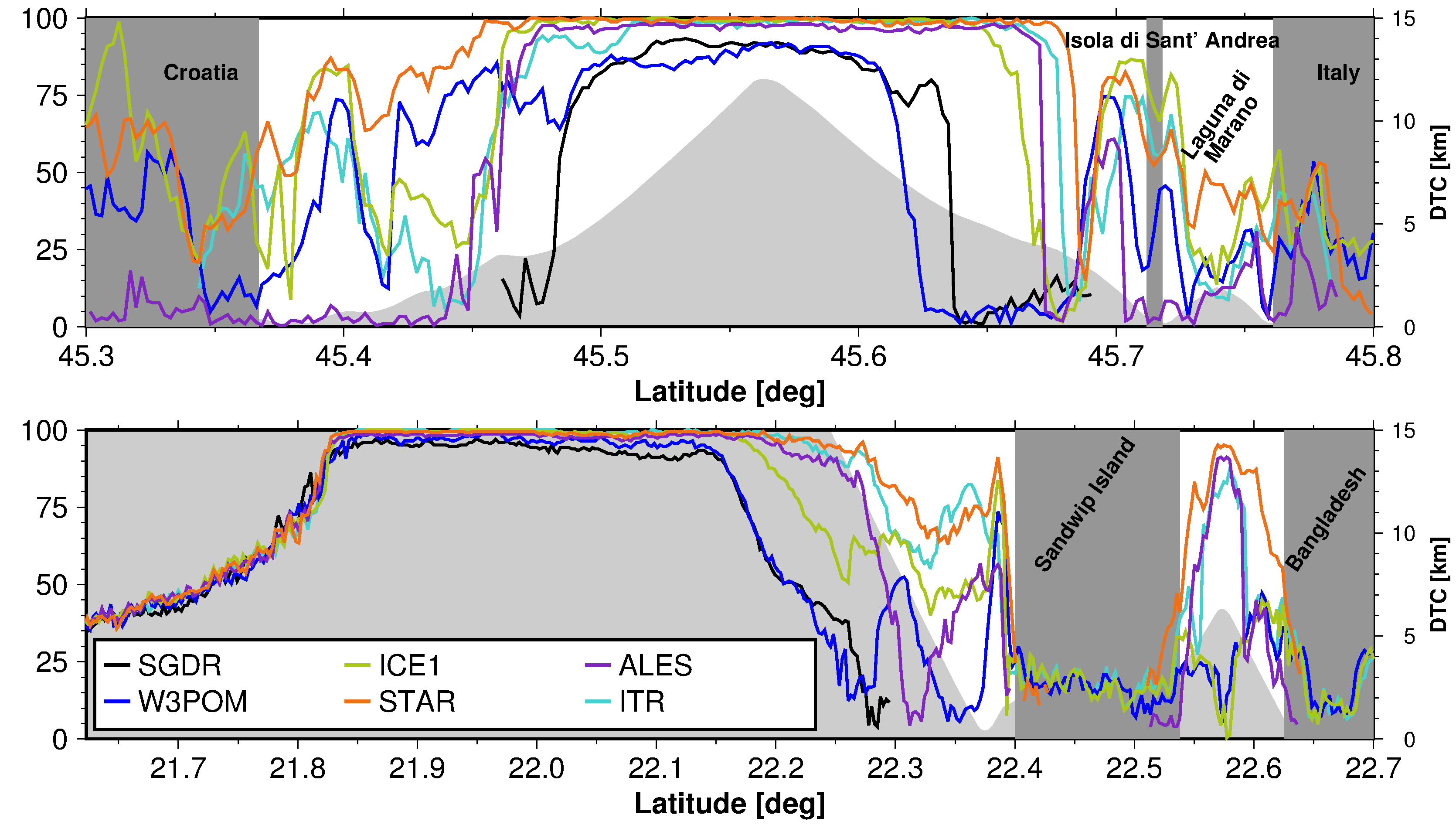}
	\caption{Percentage of cycles retained to achieve a correlation of at least $0.9$ with hourly tide gauge data from a total number of 227 available cycles. (top) Study site at the Gulf of Trieste. (bottom) Study site at the coast of Bangladesh. The distance to the nearest coastline (DTC) is provided in light gray.}
	\label{fig:numTot}
\end{figure*}

We compare the number of cycles of SSH derived from each retracker which can be utilized at each measurement location by iteratively eliminating the largest difference to the tide gauge time series from the time series above until a correlation of at least 0.9 is achieved. This enables a more direct comparison of the individual retracking methods. 
For evaluation, the number of retained cycles $N$ is plotted for each \unit[20]{Hz} along-track position in both of our study areas (Fig. \ref{fig:numTot}). 

Over the open ocean regions of our study sites in the Gulf of Trieste and off the coast of Bangladesh, we find good agreement between \threshnt{}, \alesnt{}, \itrnt{} and \STARnt{} with $95-100$\% of retained cycles. For \wtpnt{} and \sgdrnt{} we find a slightly lower level of about $90$\% of retainable cycles resulting from divergence of the parameter estimation. In addition, it is obvious that the time series of altimetry derived SSHs over the deep ocean parts of the Bay of Bengal does not agree well with time series obtained on the coastal shelf.

In the Croatian coastal area, 
all retracking methods experience a small downward peak near \unit[45.425]{$^\circ$N}. 
About \unit[4]{km} off the coast, we find a rapid drop in \alesnt{} and \itrnt{} to a level of $0-10$\% retained cycles, with \itrnt{} rising back to about $65$\% of retained cycles \unit[1]{km} off the coast. For \threshnt{} we also detect a drop to about $30$\%, followed by a rise to $80$\% at about \unit[1]{km} off the coast. About $60-80$\% of the SSHs can be retained from applying the \wtp{} model in this coastal area. \STARnt{} declines from nearly $100$\% over the open ocean to $80$\% at about \unit[$1$]{km} off the coast.   

At the Italian coast, we find nearly $100$\% of retainable cycles up to approximately \unit[3-4]{km} off the coast for \thresh{}, \itr{}, \ales{} and \STAR{}. For \threshnt{}, \itrnt{}, \alesnt{} and \STARnt{} a steep drop to less than $20$\% of retained cycles can be detected near \unit[45.675]{$^\circ$N} resulting from land influences from the small islands that separate the Laguna di Marano from the Gulf of Trieste (see also Sec. \ref{subsec:retrSSH}). The median percentage of retainable cycles for \sgdr{} and \wtp{} in this coastal area (Table \ref{tab:resultsT}) is quite low, indicating difficulties to reach convergence when the total waveform is used. The sub-waveform retrackers provide a larger number of retainable cycles as they are less affected by the land influences.
Over the central part of the Laguna di Marano, generally we find a level of $10-20$\% of retained cycles for all retrackers, except \STAR{}, which provides \STARnt{}$\approx 40\%$.

At the Bangladesh study site, \wtpnt{} and \sgdrnt{} start to decline near \unit[22.15]{$^\circ$N} towards the coast of Sandwip Island and reach a minimum level of $0-15$\% at about \unit[15]{km} off the coast. In addition, a decrease of \threshnt{}, \alesnt{}, \itrnt{} and \STARnt{} can be detected.
The \STAR{} algorithm is able to retain at least $85$\% of SSHs of the available cycles in this region. In the northern Bay of Bengal, the tidal amplitude can reach several meters which leads to significantly different land influences on the shape of the waveform between low and high tide. Consequently, this starts to affect the retrieved SSHs at about \unit[15]{km} off the coast with \alesnt{}, \itrnt{} and \STARnt{} starting to decrease towards the coast. For \alesnt{} we find a strong drop to $10$\% in front of the Sandbank region mentioned before followed by an increase to a level of \alesnt{} $\approx 50\%$ at the coast. \itrnt{} and \STARnt{} drop to a level of $70-80$\% over the sandbank area. Here, \itr{} and \STAR{} perform similar, while the overall correlations suggest a lower correlation for \itr{} in this region compared to \STAR{}, due to already removed SSHs from low tide cycles by the RANSAC algorithm as part of \STAR{}.
Over the strip of open water between Sandwip Island and the mainland of Bangladesh, \alesnt{}, \itrnt{} and \STARnt{} indicate that all three algorithms are able to retain about $85-95$\% of the available cycles in the central parts of this open water strip, while \STAR{} is able to provide a larger number of cycles towards the coastlines.

\subsubsection{Root Mean Square Difference}
\begin{figure*}[ht]
\centering
	\includegraphics[width = 0.95\textwidth]{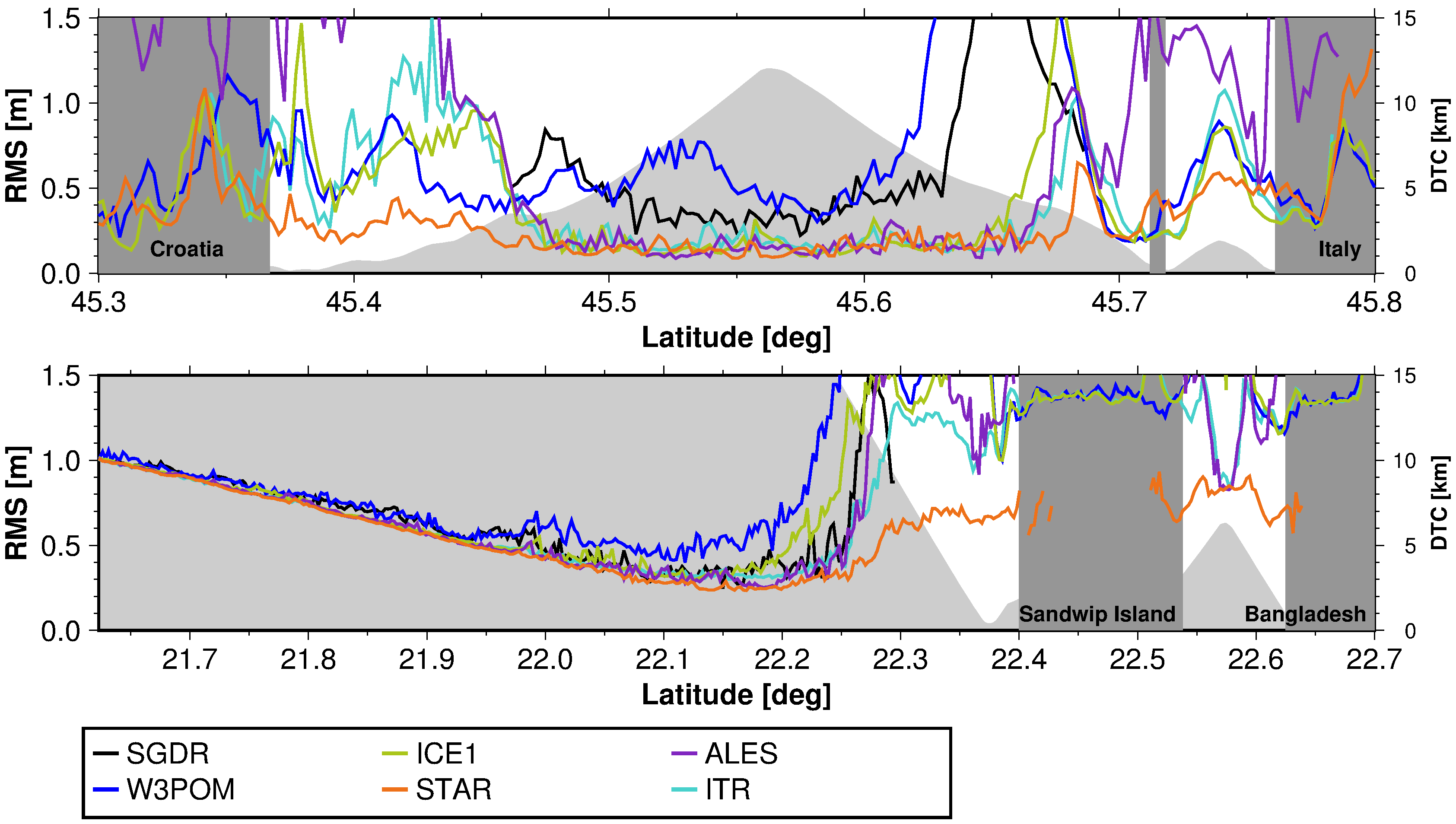}
	\caption{Root mean square difference (RMS) derived for all along-track locations after applying the selection criteria mentioned above for all available cycles (Fig. \ref{fig:numMC}) for both study sites. (top) Study site at the Gulf of Trieste. (bottom) Study site at the coast of Bangladesh. The distance to the nearest coastline (DTC) is provided in light gray.}
	\label{fig:rms}
\end{figure*}

The RMS between SSHs derived from the individual retracking methods and tide gauge data computed at each \unit[20]{Hz} along-track position provides an additional quality measure (Fig. \ref{fig:rms}). 
We apply the same criteria for computation as introduced in Sec. \ref{sec:repeat}.

Over the open ocean, we generally find results similar to the analysis of the overall correlation with \threshrms{}, \itrrms{}, \alesrms{} and \STARrms{} agreeing well at a level of \unit[0.15-0.2]{m} at the Trieste site and \unit[0.2-1.0]{m} at the Bangladesh site and \wtprms{} and \sgdrrms{} being slightly larger compared to the other methods.
 
Close to the Croatian coast, we find \threshrms{}, \itrrms{} and \alesrms{} to increase significantly with \alesrms{} reaching more than \unit[1.5]{m}. The large RMS derived for \ales{} is probably related to the small peaks in front of the leading edge. Since ALES sub-waveforms are selected from the first range gate up to the end of the leading edge, small peaks in front of the leading edge will lead to difficulties during the parameter estimation.
For \STAR{} we find a smaller increase in RMS relative to the open ocean area to \STARrms{}$\approx$ \unit[0.4]{m}. In the northern part of the Trieste site, \thresh{}, \itr{}, \ales{} and \STAR{} keep their level of RMS from the open ocean area up to a distance of \unit[4-5]{km} to the nearest coast. 

In the northern part of the Bay of Bengal (Fig. \ref{fig:rms}, bottom), the SSHs from all retracking methods become more noisy northward of about \unit[22.0]{$^\circ$N}, with a minimum RMS level of about \unit[0.25]{m}. 
Over the sandbank region, the RMS obtained from \sgdr{}, \wtp{}, \thresh{} and \ales{} show a rapid increase to at least \unit[1]{m}, while the RMS based on \STAR{} exhibits a relatively smaller increase to a level of about \unit[70]{cm} up to the coast of Sandwip Island. 
The same level is found for \STARrms{} in the small strip of open water between Sandwip Island and the mainland of Bangladesh with \alesrms{} and \itrrms{} at a similar level in the center of this region.

\subsection{Significant Wave Height and Sigma-Nought}

Besides range corrections, the use of the 3-parameter retracking model also allows to retrieve SWH and sigma-nought $\sigma^0$. The SWHs and sigma-nought are selected based on the heights chosen by the shortest-path algorithm. Theoretically it would be possible to run the Dijkstra algorithm independently on the point-clouds of SWH and sigma-nought. However, we think that the values should be consistent with the selected heights. 

No independently measured time series of wave height or wind speed is available for our two study sites. 
Therefore, we compare temporal median, RMS and the number of retained cycles of good SWH and wind speed which were obtained by individual retracking algorithms.
We further compare the obtained results to ERA-Interim model data. 
We will focus on the Bangladesh study site since no model data is available from the Gulf of Trieste area.

\subsubsection{Significant Wave Height}
\begin{figure*}[ht]
\centering
	\includegraphics[width = 0.95\textwidth]{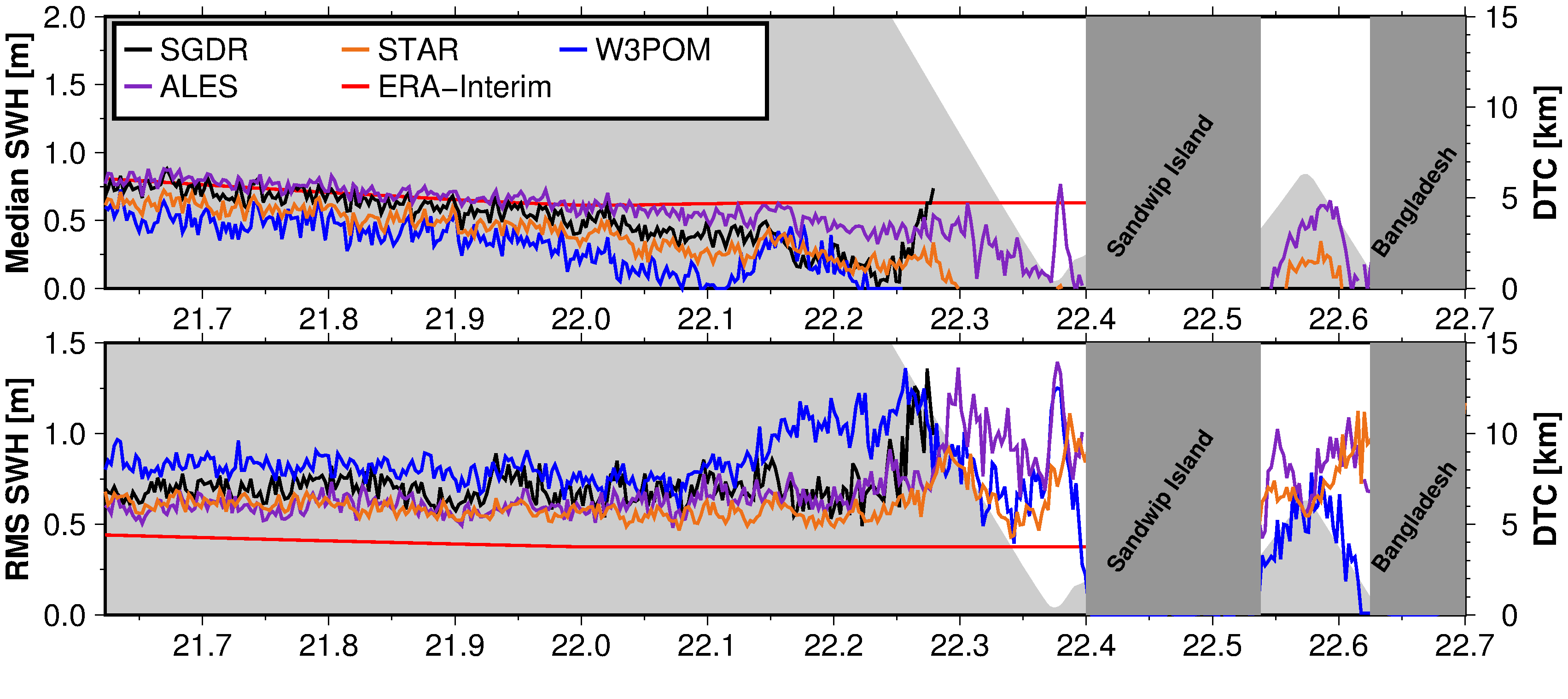}
	\caption{SWH compared to ERA-Interim model data. Top: median over all cycles, Bottom: RMS over all cycles.}
	\label{fig:swh}
\end{figure*}

Fig. \ref{fig:swh} shows the median SWH and RMS over all 239 cycles for the Bangladesh study site. 
Wave heights from \sgdr{} and \ales{} agree well with the model data over the open ocean while \STAR{} and \wtp{} are biased. 
At the same time, the median wave height values of all retrackers decline towards the coast, reaching a minimum over the sandbank area. 
On the contrary, the model data does not show this decline. 
Over open ocean, the RMS values from all retrackers agree well with \ales{} and \STAR{} providing the lowest RMS at a level of about \unit[60]{cm} while the model data suggests a level of about \unit[48]{cm}. 
At about \unit[10]{km} off the coast the RMS values from all retrackers increase due to the sandbank's influence during low tide cycles.

\begin{figure*}[ht]
\centering
	\includegraphics[width = 0.95\textwidth]{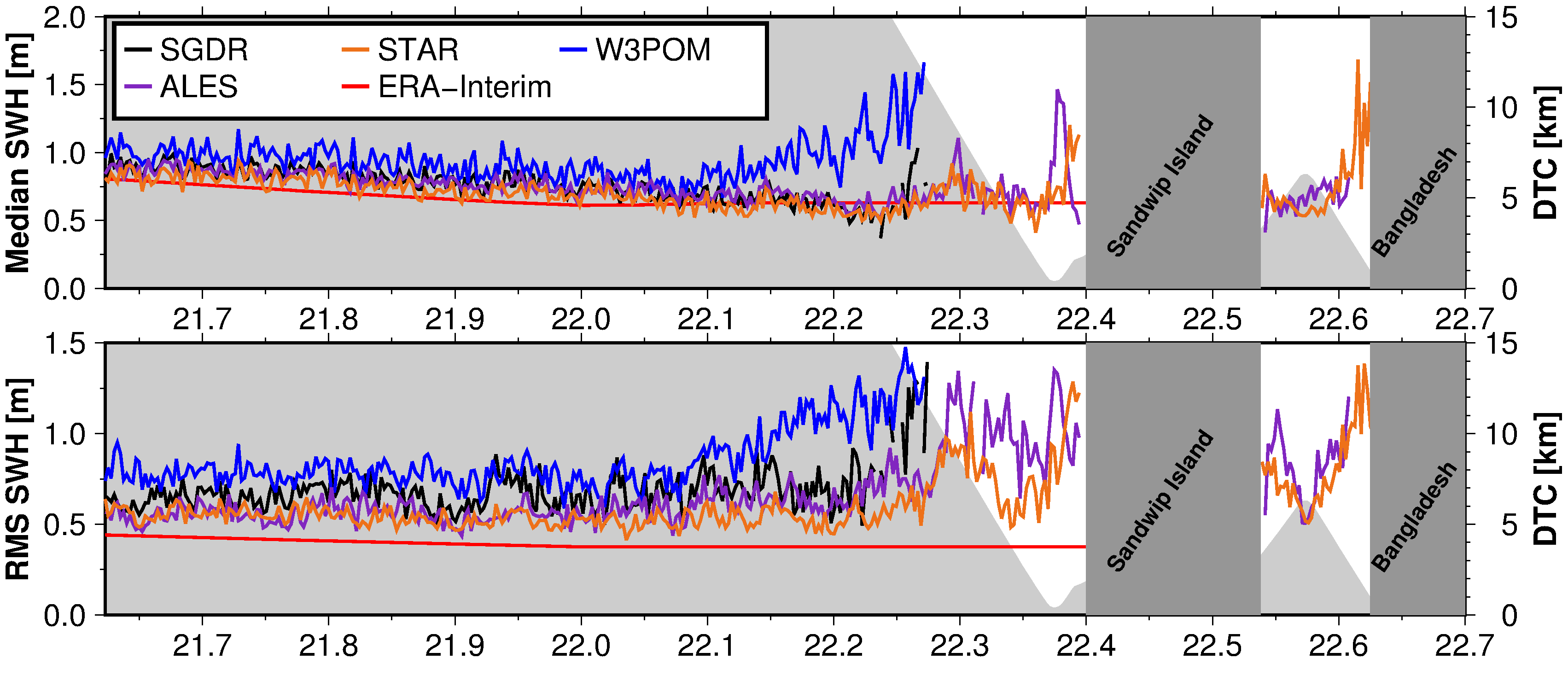}
	\caption{SWH compared to ERA-Interim model data. Top: median over all cycles, Bottom: RMS over all cycles. All zero-SWH removed.}
	\label{fig:swh2}
\end{figure*}

The low agreement between the retracked SWH and the model data, especially closer to the coast, can be explained by the estimation of the retracking parameters. 
All utilized retrackers constrain the SWH parameter to positive values since wave heights below zero are not physically meaningful. 
This sometimes leads to forcing the wave height to be zero during the estimation for individual waveforms since the leading edge, necessary for SWH parameter estimation, is only represented by a limited amount of range gates during calm sea state conditions. 
For \ales{} method the problem appears less severe since it is a two stage procedure in which SWH is estimated and averaged along-track during the first stage and kept fixed during the second stage. 
For sub-waveform retrackers such as \STAR{} the problem will occur more often in case the selected sub-waveform which contains the leading edge is relatively small. 
In case the zero wave heights are not utilized to derive the median and RMS value, \sgdr{}, \ales{} and \STAR{} agree well with the model data (Fig. \ref{fig:swh2}, Table \ref{tab:swhB}).
Over the sandbank area and the channel between Sandwip Island and the main land, \sgdr{} and \wtp{} quality decreases due to the low number of available cycles (Fig. \ref{fig:numTot}, bottom).

Zero wave heights occur for about \unit[10-15]{\%} of the cycles during calm sea state conditions at the Bangladesh study site, especially in coastal areas. Similar observations are made for the Trieste study site. 
In future, \STAR{} may be extended to a two-step procedure similar to \ales{} in order to counter these effects. 
We also applied our method to reduced SAR (RDSAR) waveforms, which are conventional altimetry like waveforms that are derived from the SAR signal during post-processing. 
The RDSAR data from Cryosat-2 data showed that the problem does not occur when zero-padding is applied during the derivation of the reduced SAR waveforms due to doubling the number of range gates available which allows better estimation of SWH (not shown here). 

\begin{table}
\caption{Median values of SWH median and RMS over different regions along the track at the Bangladesh study site. All zero-SWH removed. $M$ denotes the median and $\sigma$ the RMS.}
\centering
\begin{tabular}{lcccccc}
\hline
\textbf{Retracker} & \multicolumn{2}{c}{Open Ocean} & \multicolumn{2}{c}{Sandbank} & \multicolumn{2}{c}{Channel} \\ 
 & $M$ [m] & $\sigma$ [m] & $M$ [m] & $\sigma$ [m] & $M$ [m] & $\sigma$ [m] \\ \hline
SGDR & 0.94 & 0.69 & 0.59 & 0.80 & - & - \\
W3POM & 1.06 & 0.80 & 1.13 & 1.20 & - & - \\
ALES & 0.78 & 0.57 & 0.66 & 0.86 & 0.67 & 0.82 \\
STAR & 0.89 & 0.59 & 0.62 & 0.68 & 0.65 & 0.71 \\ \hline
ERA-I & 0.84 & 0.45 & 0.63 & 0.38 & - & - \\
 \hline

\end{tabular}
\label{tab:swhB}
\end{table}

\subsubsection{Wind Speed}

\begin{figure*}[ht]
\centering
	\includegraphics[width = 0.95\textwidth]{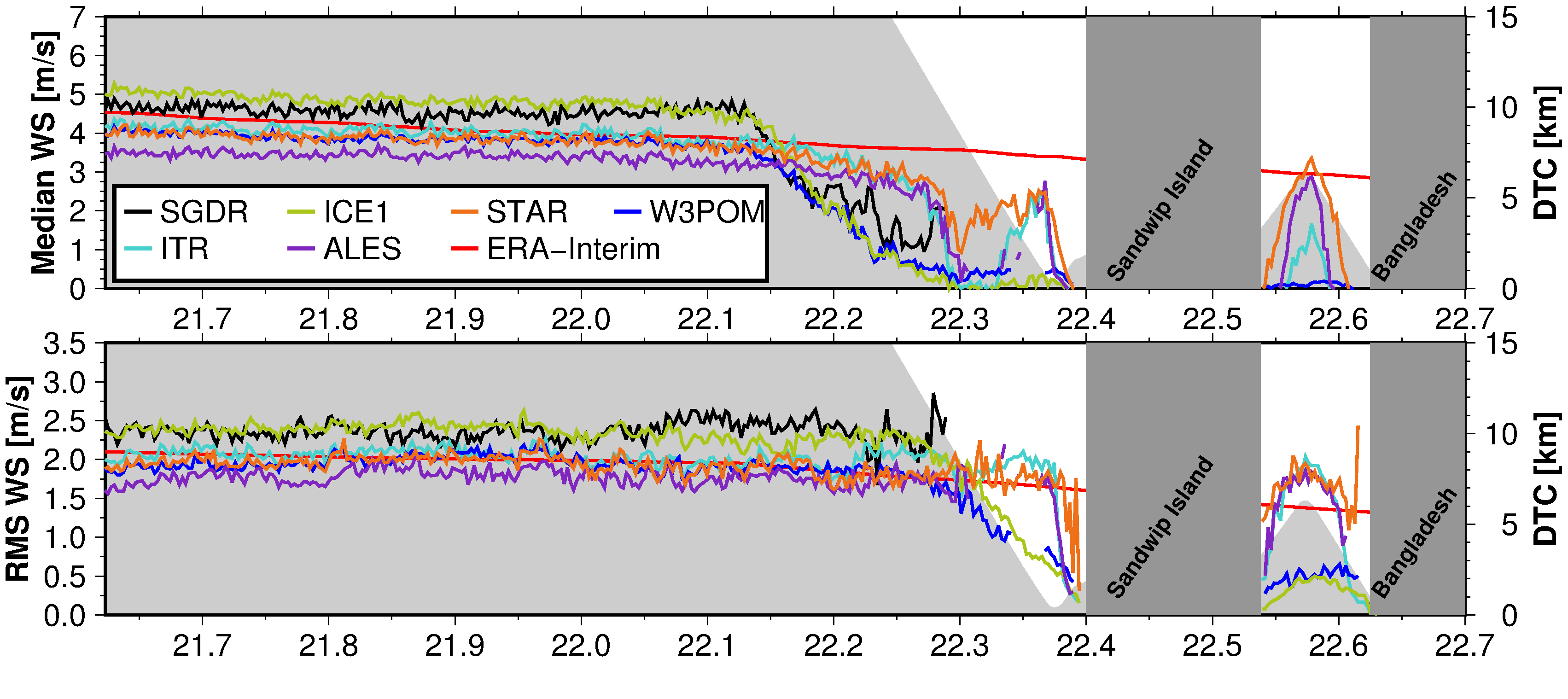}
	\caption{Wind Speed compared to ERA-Interim model data. Top: median over all cycles, Bottom: RMS over all cycles.}
	\label{fig:wind}
\end{figure*}

Wind speed has been derived from the estimated amplitudes utilizing the 2-parameter model by \citet{Gourrion2002} as described in \citet{Aviso2015}. The model input consists of the retracked sigma-nought and SWH. Here, we utilized only non-zero SWH in order to keep the wind speed estimate unbiased from problems in SWH estimation. Similar to SWH, we display the median and RMS for each retracker and from the ERA-Interim model data at each measurement position.

Over the open ocean, median wind speed and RMS from \sgdr{}, \wtp{}, \thresh{}, \itr{} and \STAR{} agree well with the wind speed derived from ERA-Interim model data (Fig. \ref{fig:wind}, Table \ref{tab:windB}). However, \sgdr{} and \thresh{} are slightly biased above the model data, while \wtp{}, \itr{} and \STAR{} are slightly biased below the model data. \ales{} shows a stronger bias with respect to the model data. From about $22.15^\circ$N towards the coast the waveforms start to get influenced by land returns leading to peaks moving along the trailing edge towards the leading edge. Retrackers that utilize the full waveform, such as \sgdr{}, \wtp{} and \thresh{}, return amplitude estimates which are biased by the peak influence and thus do no longer provide reliable wind speed. Sub-waveform method, such as \itr{}, \ales{} and \STAR{} also show a slight decline in wind speed, and over the sandbank area where one finds mostly specular peak waveforms, the derived wind speed, especially during low tide, is no longer meaningful. In the center of the channel between Sandwip Island and the Bangladesh main land only \ales{} and \STAR{} are able to provide meaningful wind speed. 

The reason for the bias in wind speed between \sgdr{} and \wtp{} and \STAR{} is due to the chosen estimation model. For the \sgdr{} data a MLE4 estimation method is employed utilizing a 4 parameter model \citep{Amarouche2004}. The fourth parameter is the off-nadir angle that can be derived from the slope of the trailing edge and which is assumed to be zero for the 3-parameter models. The off-nadir angle influences the estimation of the amplitude \citep[][Eq. 8]{Amarouche2004} which will lead to a general bias in amplitude in case the off-nadir angle is different from zero. Here, the derived off-nadir angle is slightly positive which leads to a smaller amplitude and consequently higher estimated wind speed. Even small changes in sigma-nought have significant influence on the derived wind speed. However, for sub-waveform methods, such as \STAR{}, it is not feasible to try and estimate the off-nadir angle since the sub-waveforms do not include enough range gates from the trailing edge for a reliable estimation.

\begin{table}
\caption{Median values of wind speed median and RMS over different regions along the track at the Bangladesh study site. $M$ denotes the median and $\sigma$ the RMS.}
\centering
\begin{tabular}{lcccccc}
\hline
\textbf{Retracker} & \multicolumn{2}{c}{Open Ocean} & \multicolumn{2}{c}{Sandbank} & \multicolumn{2}{c}{Channel} \\ 
 & $M$ [m/s] & $\sigma$ [m/s] & $M$ [m/s] & $\sigma$ [m/s] & $M$ [m/s] & $\sigma$ [m/s] \\ \hline
SGDR & 4.57 & 2.30 & 2.31 & 1.86 & - & - \\
W3POM & 3.85 & 1.89 & 0.49 & 1.63 & 0.10 & 0.50 \\
ICE1 & 4.85 & 2.33 & 0.23 & 1.90 & -0.15 & 0.38 \\
ITR & 4.06 & 2.02 & 1.76 & 1.98 & 0.01 & 1.61 \\
ALES & 3.35 & 1.72 & 2.42 & 1.68 & 1.02 & 1.68 \\
STAR & 3.86 & 1.91 & 2.43 & 1.77 & 2.31 & 1.70 \\ \hline
ERA-I & 4.60 & 2.12 & 3.58 & 1.73 & 2.94 & 1.37 \\
 \hline

\end{tabular}
\label{tab:windB}
\end{table}

\subsection{Application to Jason-1 and Envisat Data}

\begin{figure*}[ht]
\centering
	\includegraphics[width = 0.95\textwidth]{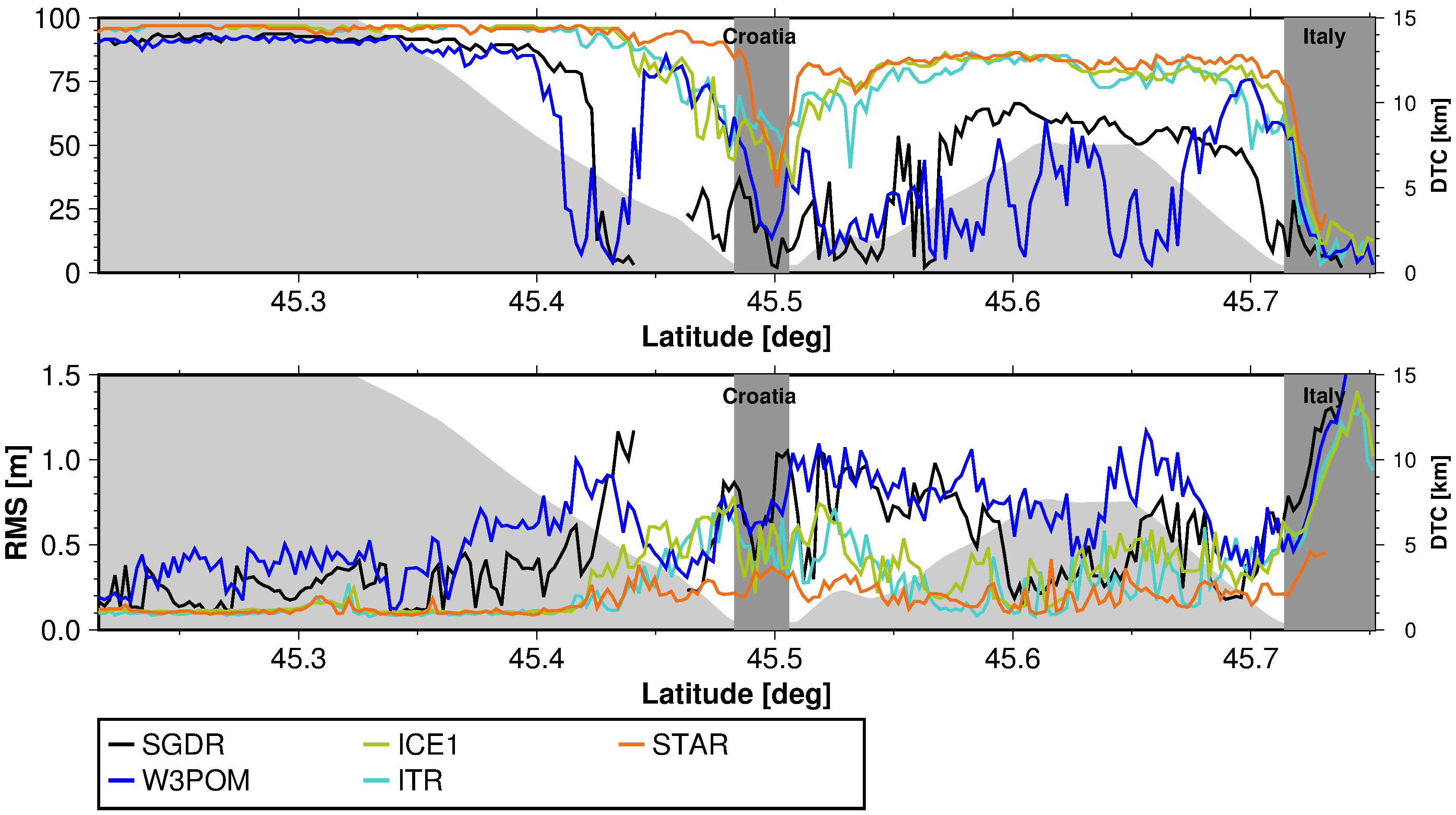}
	\caption{(top) Percentage of cycles retained to achieve a correlation of at least $0.9$ with the hourly tide gauge data from a total number of 95 available cycles of the Jason-1 interleaved mission. (bottom) Root mean square difference (RMS) derived for all along-track locations. The distance to the nearest coastline (DTC) is provided in light gray.}
	\label{fig:j1}
\end{figure*}

\begin{figure*}[ht]
\centering
	\includegraphics[width = 0.95\textwidth]{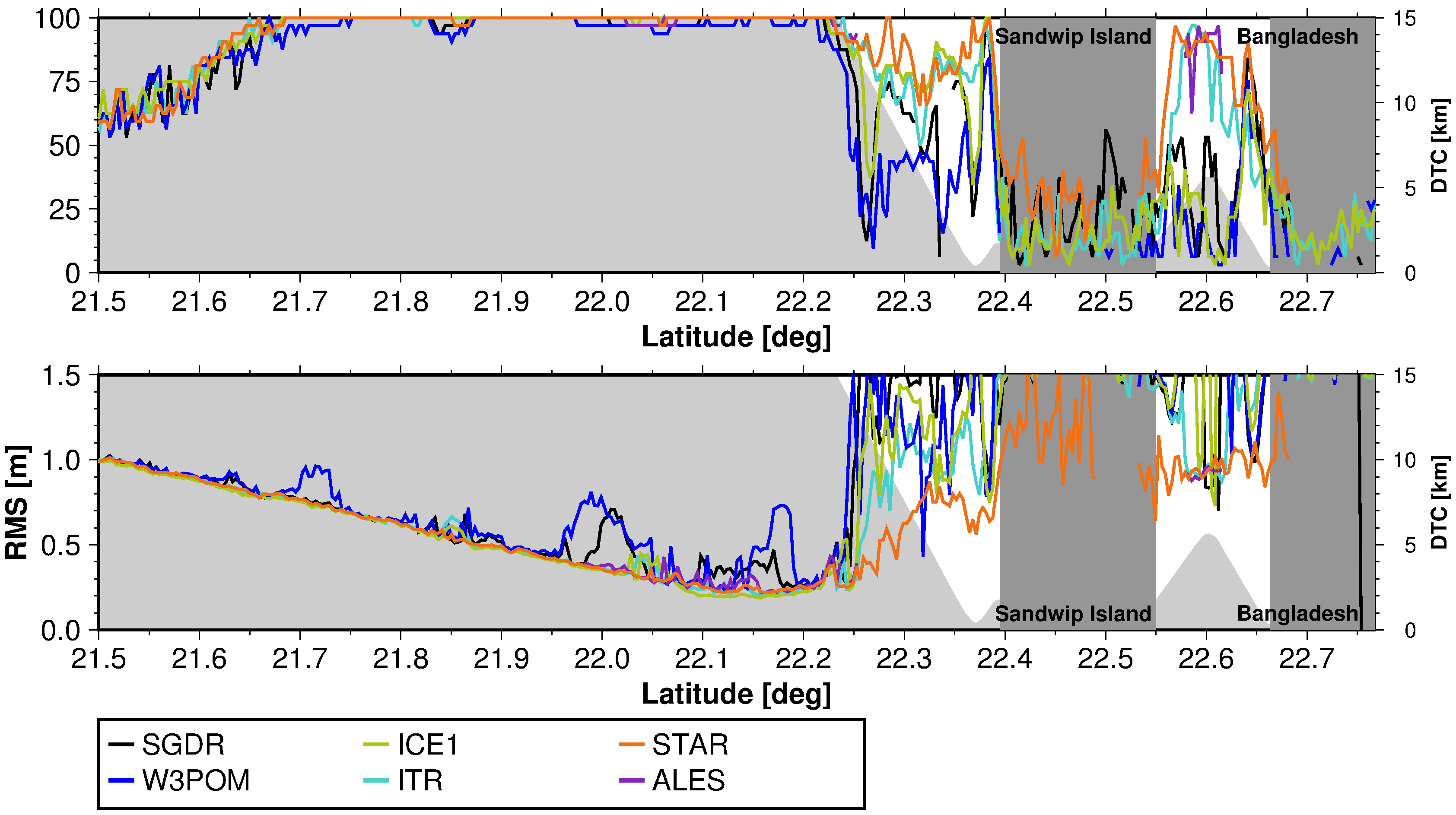}
	\caption{(top) Percentage of cycles retained to achieve a correlation of at least $0.9$ with the hourly tide gauge data from a total number of 32 available cycles of the Envisat mission. (bottom) Root mean square differences (RMS) derived for all along-track locations. The distance to the nearest coastline (DTC) is provided in light gray.}
	\label{fig:envi}
\end{figure*}

We investigate the application of \STAR{} to altimetry data from the Jason-1 interleaved period to the Trieste site (Fig. \ref{fig:study}(a)) and to Envisat data for the Banlgadesh site (Fig. \ref{fig:study}(b)). 
However for Jason-1 interleaved, there is no \ales{}-data available for comparison. 
Here, we focus on the number of cycles retained to reach a correlation of $>0.9$ (Fig. \ref{fig:j1} and \ref{fig:envi}, top), as well as the RMS at each along-track position (Fig. \ref{fig:j1} and \ref{fig:envi}, bottom). For Jason-1, we found 95 cycles that overlapped with the available time period for the tide gauge data at the Trieste station, while we found only 32 cycles of Envisat data that overlapped with the tide gauge period available for the Chittagong station.

Utilizing data from the Jason-1 mission over the open ocean at the Trieste study site, \thresh{}, \itr{} and \STAR{} show more retained cycles, as well as a significantly smaller RMS compared to \sgdr{}- and \wtp{}-based SSHs. 
In the coastal area over about the last \unit[5]{km} in front of the Croatian coast, \STAR{} derived SSHs fit well to the tide gauge data (Fig. \ref{fig:j1}). 
After crossing parts of the Croatian mainland, the track transitions back to the ocean and we find \thresh{}, \itr{} and \STAR{} retaining significantly more cycles compared to \sgdr{}- and \wtp{}. The general level of retained cycles is lower due to signal losses that happened between the coasts of Croatia and Italy during nine of the Jason-1 interleaved cycles. At the Italian coast, all \itr{}, \thresh{} and \STAR{} are able to maintain a high quality of SSHs in the coastal area with which decreases only slightly over the last \unit[1-2]{km} off the coast. 

Additionally, we have utilized Envisat track 416 for our experiments, which crosses our Trieste study site from the north-east to the south-west. However, the overlapping period with the available tide gauge data was less than one year and deriving any correlations or RMS from this short period containing only $10$ cycles of the $35$-day Envisat repeat orbit would not be meaningful. Nonetheless, SSHs derived from these 10 cycles of Envisat data compared well to the available tide gauge data (not shown here).

For the study site at the coast of Bangladesh, we found an overlap period of more than $3$ years between $32$ cycles (2007 - 2010) of  Envisat data  and the Chittagong tide gauge which allows to derive meaningful correlations and RMS differences. 
Over the open ocean area, the number of retained cycles and the RMS agree well between all the considered retracking methods with \wtp{} bases SSHs being slightly more noisy (Fig. \ref{fig:envi}). 
About \unit[15]{km} before the Envisat track reaches Sandwip Island, the differences between the individual retracking methods become evident. 
Ratios of retained cycles for \wtp{} and \sgdr{} show a rapid drop to less than $25\%$ over the sandbank area and a gradual increase towards the coast of Sandwip Island to a level of $35\%$ and $60\%$, respectively. 
This behavior is combined with a strong increase in RMS differences which increase to more than \unit[1]{m}. Similar behavior can be observed for \thresh{} and \itr{}, which both show a smaller drop in the number of retained cycles over the last \unit[15]{km} towards the coast to a level of about $75\%$, but their corresponding RMS increases significantly to values greater than \unit[1]{m}. For \ales{} there is no data available over the last \unit[15]{km} in front of Sandwip Island. \STAR{} shows the smallest drop in number of retained cycles and recovers back to a level of more than $90\%$ close to Sandwip Island. 
Over the small strip of open water between Sandwip Island and the mainland of Bangladesh, the coastal retrackers \itr{}, \ales{} and \STAR{} retain about $90\%$ of the available cycles providing SSHs that fit well to the tide gauge data. 
\sgdr{}, \wtp{} and \thresh{} are not able to provide more than $0-25$\% of retained cycles and exhibit significantly higher RMSs in this region.


\section{Conclusion}
\label{sec:conclusion}

A novel method for analyzing altimetric waveforms and deriving sea surface heights, SWH and sigma-nought has been suggested.
The proposed technique partitions the total waveform into individual sub-waveforms which can be analyzed in combination with existing retracking models. 
The sea surface heights provided by STAR were found to be of at least the same quality or better compared to existing conventional and coastal retracking methods over the open ocean, as well as in coastal regions. In addition, correlations with tide gauge data revealed generally more usable cycles close to the coast in combination with lower root mean square differences compared to existing methods.
Of course, depending on the retracking model that is combined with the derived sub-waveforms, it is possible to derive significant wave height and backscatter in the same way. 

Comparison of the derived sub-waveforms with the \citet{Hwang2006} method reveal good correspondence between identified parts of the waveform. 
We found the influence of the random component of STAR on the SSH results to be at a level of less than \unit[5]{cm} over the open ocean and at about \unit[20]{cm} in coastal regions. This is in range with modifications that can be applied to conventional retracking algorithms including biases between different retracking methods, different weighting schemes or varying estimators; these effects are considered to be in the order of a few centimeters.
Sea surface heights derived from the STAR algorithm have been extensively validated for Jason-2 data and compared to five independent available retracking methods, as well as hourly in-situ tide gauge measurements.
At the study sites in the Gulf of Trieste, Italy and off the coast of Bangladesh, we found varying surface conditions including (deep) open ocean and shallow coastal waters, temporally submerged sandbanks and transition zones between river estuaries and the ocean. 
Consequently, deriving five partitionings of the total waveform enabled the STAR algorithm to handle a larger variety of waveform shapes compared to existing coastal retracking algorithms.
Examination of estimated SWH and wind speed revealed good agreement to other retracking algorithms as well as model data from ERA-Interim.
Furthermore, we applied STAR to altimetry data from the Jason-1 interleaved period, as well as Envisat , which also resulted in significant improvements in the quality of coastal sea surface heights.

We are confident that the STAR method will enable a wide range of further studies, including a more comprehensive validation of significant wave height and sigma-nought. In this context, we will also consider an improved selection of final retacking results by considering SWH, sigma-nought etc. instead of only utilizing the SSHs. 
In addition, the algorithm can be improved further by tuning the a-priori hyperparameters that control, \eg{}, the resolution of the sub-waveform, as well as extending the Dijkstra algorithm to reduce the impact of potentially non-optimal sea surface heights of neighboring measurement locations. 
The derived sub-waveforms will be combined with different available waveform models in order to adapt and extend the concept to other regions, such as rivers and lakes. 
One might also consider a slightly different approach by using the partitioning into sub-waveform to derive weighting schemes for retracking of the whole waveform. 



\section*{Acknowledgements}
We acknowledge funding through the BanD-AID (KU 1207/19-1) project.
We also thank the associate editor and three anonymous reviewers for their constructive comments which helped us to significantly improve the manuscript.

\bibliographystyle{model5-names}

\end{document}